\documentclass[journal]{IEEEtran}
\usepackage[colorlinks,
            linkcolor=blue,
            anchorcolor=blue,
            citecolor=blue]{hyperref}

\usepackage{cite}
\usepackage{amsmath,amssymb,amsfonts}
\usepackage{graphicx}
\usepackage{textcomp}
\usepackage{xcolor}

\usepackage[numbers]{natbib}
\usepackage{listings}
\usepackage{amsmath, amssymb}
\usepackage{makecell}
\usepackage{multirow}
\usepackage{booktabs}
\usepackage{subfigure}
\usepackage{xcolor}
\usepackage{listings}
\usepackage{algorithm}
\usepackage{algpseudocode}
\usepackage{graphics}
\usepackage{epsfig}
\usepackage{array}
\usepackage{autobreak}
\usepackage{color}
\usepackage{threeparttable}
\usepackage{soul}
\soulregister\cite7
\soulregister\url7

\ifCLASSINFOpdf
\else
\fi

\begin{document}
%
\title{Inter-BIN: Interaction-based Cross-architecture IoT Binary Similarity Comparison\\ }
%
%
%

%
%

\markboth{IEEE Internet of Things Journal}%
{Shell \MakeLowercase{\textit{et al.}}: Bare Demo of IEEEtran.cls for IEEE Journals}
%



\author{Qige Song,
		Yongzheng Zhang,~\IEEEmembership{Member,~IEEE},
        Binglai Wang,
        Yige Chen


\thanks{This work was supported in part by the National Key Research and Development Program of China under Grant 2018YFB0804704, in part by the National Natural Science of China under Grant U1736218, and in part by the Beijing Municipal Science and Technology Commission under Grant Z191100007119005. (\emph{Corresponding author: Yongzheng Zhang.})}


\thanks{The authors are with the Institute of Information Engineering, Chinese Academy of Sciences, Beijing 100093, China, and also with the School of Cyber Security, University of Chinese Academy of Sciences, Beijing 100049, China (e-mail: songqige@iie.ac.cn; zhangyz@cacts.cn; wangbinglai@iie.ac.cn; chenyige@iie.ac.cn).}}

\maketitle


\begin{abstract}
The big wave of Internet of Things (IoT) malware reflects the fragility of the current IoT ecosystem. Research has found that IoT malware can spread quickly on devices of different processer architectures, which leads our attention to cross-architecture binary similarity comparison technology. The goal of binary similarity comparison is to determine whether the semantics of two binary snippets is similar. Existing learning-based approaches usually learn the representations of binary code snippets individually and perform similarity matching based on the distance metric, without considering inter-binary semantic interactions. Moreover, they often rely on the large-scale external code corpus for instruction embeddings pre-training, which is heavyweight and easy to suffer the out-of-vocabulary (OOV) problem. In this paper, we propose an interaction-based cross-architecture IoT binary similarity comparison system, \texttt{Inter}-\texttt{BIN}. Our key insight is to introduce interaction between instruction sequences by co-attention mechanism, which can flexibly perform soft alignment of semantically related instructions from different architectures. And we design a lightweight multi-feature fusion-based instruction embedding method, which can avoid the heavy workload and the OOV problem of previous approaches. Extensive experiments show that \texttt{Inter}-\texttt{BIN} can significantly outperform state-of-the-art approaches on cross-architecture binary similarity comparison tasks of different input granularities. Furthermore, we present an IoT malware function matching dataset from real network environments, \emph{CrossMal}, containing 1,878,437 cross-architecture reuse function pairs. Experimental results on \emph{CrossMal} prove that \texttt{Inter}-\texttt{BIN} is practical and scalable on real-world binary similarity comparison collections.
 
\end{abstract}


%
\IEEEpeerreviewmaketitle

\section{Introduction}
\subsection{Background and Motivation}
\IEEEPARstart{B}{inary} code similarity comparison (or matching) aims to detect whether the semantics of two given pieces of binary code are similar or not. It is a significant issue in software security analysis and has a wide range of application scenarios, such as vulnerability detection, malware analysis, software plagiarism detection, and code authorship verification. Recently, security issues in new application scenarios have led us to pay attention to cross-architecture binary similarity analysis, including the big wave of Internet of Things (IoT) malware \cite{costin2018iot} \cite{alasmary2019analyzing}. As the extension and development of the Internet, IoT technology has been widely used in various industries, such as intelligent transportation, smart medical care, and industrial control systems, which greatly facilitate our lives. The fast-developing IoT system introduces a wide variety of devices, and the platforms and functions of the devices are highly heterogeneous. It is estimated that by 2025, there will be more than 30 billion IoT connections worldwide, with nearly four IoT devices per person \footnote{\url{https://iot-analytics.com/iot-2020-in-review/}}. Due to the rapid increase in demand for IoT devices and applications, developers are focused on quickly implementing the core functions of their products and launching them on the market, while the security issues of the IoT environment have not received sufficient attention. Many IoT devices and software have not yet reached the current security standards and suffer vulnerabilities and weaknesses, making them a new hot target of malware developers. In 2016, the malware family \texttt{Mirai} infected hundreds of thousands of IoT devices and operated them to launch large-scale DDoS attacks, causing massive damage and reflecting the fragility of the current IoT ecosystem\cite{IEEEexample:antonakakis2017understanding}.

Compared with the malware families of desktops and the Android platform that have been extensively studied, the analysis for the IoT malware is currently not comprehensive and systematic enough. Previous research deploys IoT honeypots to simulate fragile IoT devices and capture malware instances for in-depth analysis. The designers of \texttt{IoTPOT} \cite{pa2016iotpot} and \texttt{IoTCMal} \cite{wang2020iotcmal} found that some malware families evolved rapidly in a short period and quickly reused and disseminated on a large number of devices of diverse CPU architectures. Based on this characteristic, a practical cross-architecture binary code matching solution can help efficiently discover malware targeting the IoT devices of different architectures.

\begin{figure}[!t]
	\centering
	\includegraphics[width=8cm]{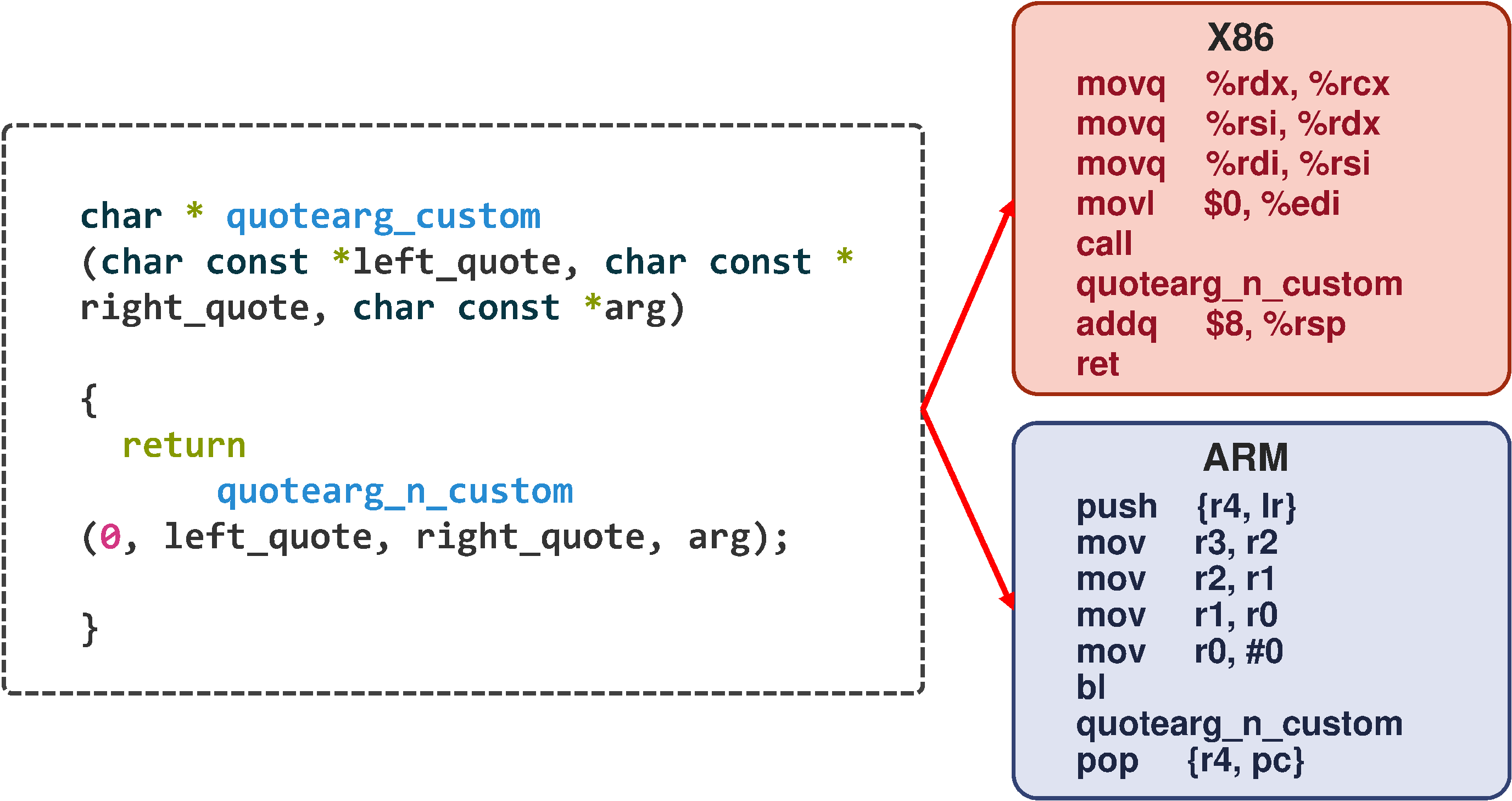}
	\caption{Assembly code on two architectures complied from the same source code snippet} 		
	\label{intro_cmp}
\end{figure}

\begin{figure}[!t]
	\centering
	\includegraphics[width=6cm]{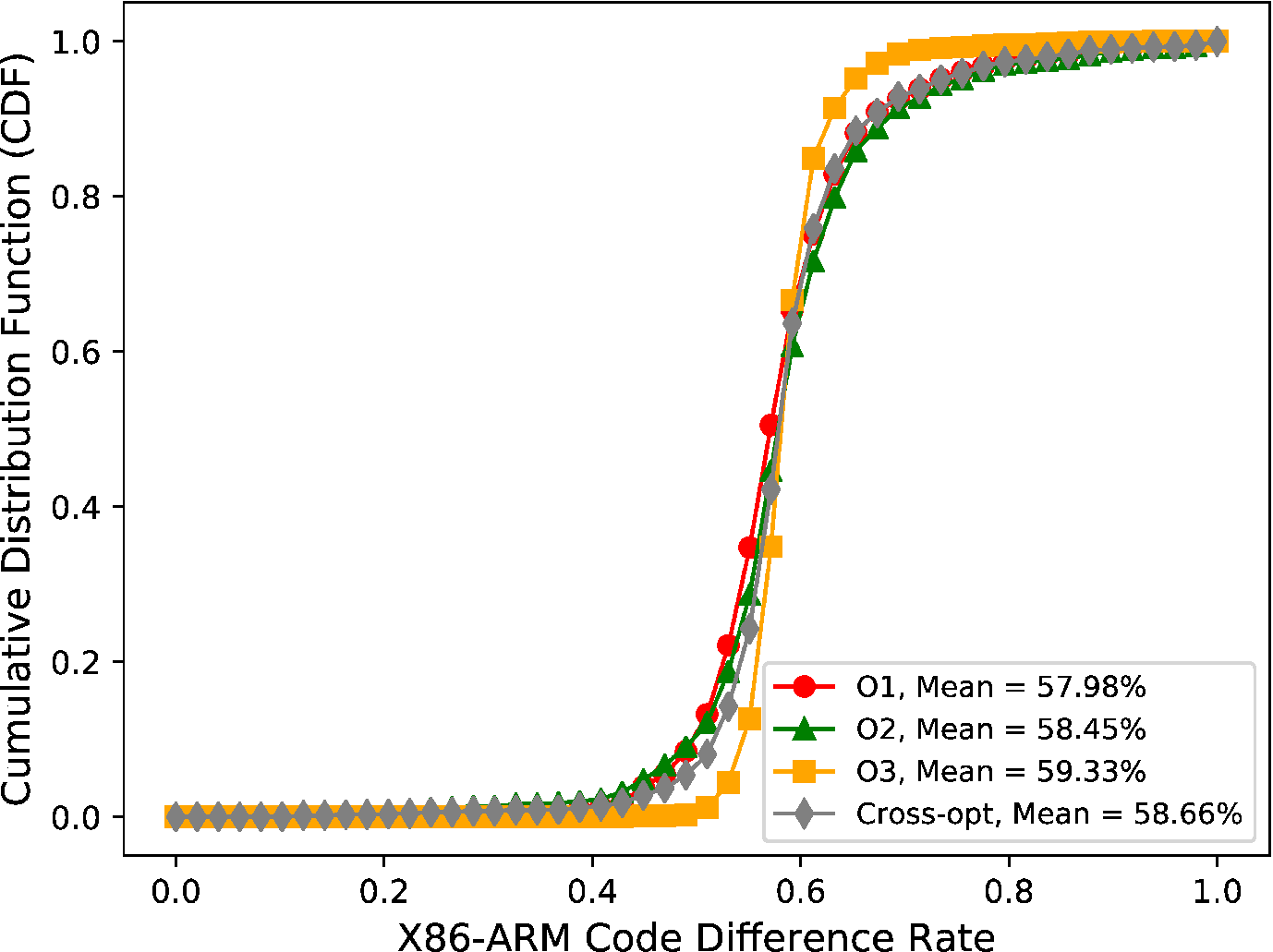}
	\caption{The CDF of cross-architecture code difference rate.}
	\label{intro_cdf}
\end{figure}

Cross-architecture binary similarity comparison is non-trivial because different architectures have separate instruction sets with different mnemonics, CPU registers, calling conventions, and memory access strategies \cite{haq2019survey} \cite{pewny2015cross}. \figurename \ref{intro_cmp} shows the assembly code compiled from the same code snippet separately on x86 and ARM architectures, and the generated instruction sequences look completely different. We compile several popular Linux packages on x86 and ARM for quantitative statistical analysis. \figurename  \ref {intro_cdf} is the cumulative distribution function (CDF) of assembly code difference rate complied from the same source code. Under the same compiler and optimization level, the cross-architecture function pairs have an average of 58.34\% code differences. Therefore, it is difficult to determine whether a pair of cross-architecture binaries are similar or not by matching the lexical and syntax characteristics of their instruction sequence.

Moreover, we consider that a binary comparison approach should be able to support different input granularities flexibly. Malware developers may reuse malicious modules and functions of existing instances, or deconstruct the benign program's functions and only add small code pieces to inject malicious behavior. Code copyists may splice and rearrange the original procedures or functions to hide their plagiarism intentions. In these cases, coarse-grained binary semantic similarity comparison like whole program level is difficult to discover suspicious behavior or detect software plagiarism accurately.

Motivated by the practical value of the cross-architecture binary similarity comparison problem, our goal is to implement a multi-granularity universal binary matching framework, which can match semantically similar binaries from different architectures, and provide efficient and scalable defense solutions for cross-architecture reuse IoT malware threats.

\subsection{Limitation of Prior Art}
Traditional binary similarity analysis approaches compare the functions of the code by monitoring their runtime behavior \cite{egele2014blanket} \cite{ming2017binsim} or extracting syntax features of binary sequences \cite{santos2010idea} \cite{xu2017spain}. Recently, researchers propose learning-based approaches to improve the binary comparison accuracy and scalability in cross-architecture scenarios. A binary file, after disassembled, is represented as an \emph{assembly instruction sequence}, and each instruction is composed of an opcode and zero to several operands. \texttt{INNEREYE} \cite{zuo2018neural} is a state-of-the-art approach modeling binary fragments by RNN layers and using cosine similarity as the semantic comparison metric. The disadvantage of \texttt{INNEREYE} is that the encoding process of the two input instruction sequences is trained individually, without considering the correlation of semantically similar instructions from different architectures. Redmond et al. \cite{redmond2018cross} associate cross-architecture instruction pairs through linear mapping of their position indexes within the sequences. However, this \emph{hard alignment} way is not accurate because the number and order of instructions in a pair of similar binary snippets from separate  platforms may be very different, as shown in figure \ref{intro_cmp}. Faced with the above challenges, we introduce an \emph{inter-sequence interaction scheme} that can realize the precise \emph{soft alignment} of cross-architecture semantically related instruction pairs. 

Another limitation of previous learning-based binary comparison work is that they often rely on the large-scale external code corpus to pre-train instruction \emph{embeddings} (i.e., numerical vectors) \cite{zuo2018neural} \cite{redmond2018cross} \cite{duan2020deepbindiff} \cite{massarelli2019SAFE}. The collection and processing of the code corpus are labor-intensive, and this method easily suffers the \emph{out-of-vocabulary} (OOV) problem. Specifically, the OOV problem occurs because the external binary corpus collected for pre-training cannot comprehensively include disassembly instructions in various compilation environments. Therefore, there are instructions of the input binary pieces that are not appeared in the pre-trained instruction vocabulary, and we cannot obtain their corresponding embeddings for the subsequent binary semantic comparison process. OOV is a well-known problem in the natural languages processing field. With IoT binary similarity comparison as the target domain, this problem will be severe since the binaries are likely to be compiled from various platforms and compile settings, so it is unrealistic to train a comprehensive vocabulary covering all possible dissambly instructions.

To solve this problem, we consider modeling instructions based on the fusion of multiple lightweight features. We build a character (char) dictionary table for cross-architecture instructions. The character-level (char-level) processing way can effectively reduce the vocabulary scale. \texttt{INNEREYE}'s vocabulary size reached 49,760 under one compiler setting and two architectures, while the size of our char table can be controlled to dozens, and we almost ensure that OOV will not occur. Then we perform opcode embedding and operands attributes extraction to learn more precise instruction semantic information. We fuse the char-level spatial instruction features and the semantic features to generate meaningful instruction representations without pre-training.

\subsection{Proposed Approach}
In this paper, we propose an interaction-based cross-architecture IoT binary similarity comparison approach, \texttt{Inter}-\texttt{BIN}. Given assembly instruction sequences from different architectures, we devise a multi-feature fusion method to fully extract the semantic information of instructions without relying on the context information provided by the external code collections. Specifically, we first extract instructions' spatial features by a character (char) embedding layer and a 1-D convolutional layer, which can generate meaningful char-level n-gram patterns. Then we extract statistical attributes of the preprocessed instruction and conduct an opcode embedding layer and operands feature mapping layer to further characterize the instructions. The captured semantic features of different views are concatenated as the final instruction embedding vectors. We use Bi-directional Long Short-Term Memory (Bi-LSTM) encoders to model the context information and generate representations of the two instruction sequences. To realize the information interaction between the instruction sequences, we perform automatic soft alignment of cross-architecture instructions by a co-attention mechanism. It will assign high weights to the instruction pair with high semantic correlation. Finally, we concatenate the sequence representations enhanced by interaction and use a fully-connected layer to determine their similarity comparison result. 

The key insight of our method is to introduce inter-sequence interaction scheme into the cross-architecture IoT binary similarity matching problem for finding similar functional instruction pairs with different lexical and syntax expressions. To the best of our knowledge, this is the first work that uses a \emph{deep neural network} with an \emph{interaction mechanism} for cross-architecture binary comparison. Moreover, the instruction representation module of \texttt{Inter}-\texttt{BIN} fuses multiple lightweight instruction features without relying on the large-scale external code corpus for instruction pre-training. It can significantly reduce the workload, and can adaptively replace the instruction embedding module of the existing binary similarity comparison approaches, alleviating their performance loss when suffering serious OOV problems on the evaluation dataset. 

\begin{figure*}[!t]
	\centering
	\includegraphics[width=12.5cm]{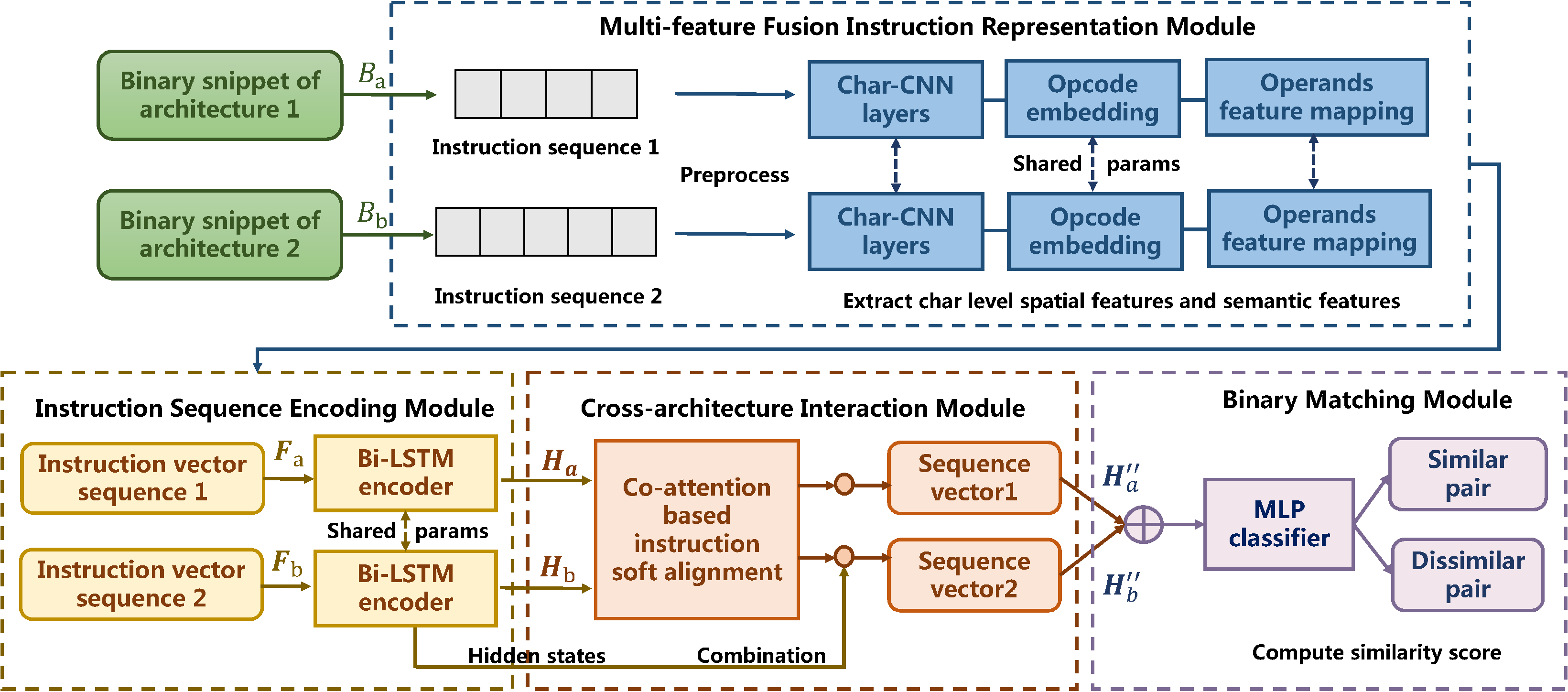}
	\caption{The architecture of \texttt{Inter}-\texttt{BIN}.}
	\label{workflow}
	\vspace*{-0.8\baselineskip}
\end{figure*}

\subsection{Key Contributions}
We summarize our major contributions as follows:
\begin{itemize}
	\item We propose an IoT binary similarity comparison approach introducing semantic interaction between different architectures' instruction sequences. It performs automatic flexible soft alignment of instruction pairs and models their functional correlations, which significantly improves the cross-architecture binary matching accuracy. And we devise a multi-feature fusion-based instruction embedding method, which can avoid the heavy workload and the OOV problem of commonly used instruction pre-training approaches.
	
    \item We implement our solution as an interaction-based cross-architecture IoT binary similarity comparison system, \texttt{Inter}-\texttt{BIN}. \texttt{Inter}-\texttt{BIN} can receive binary snippets of different granularities, perform end-to-end binary representation and inter-binary interaction process, and accurately matching input pairs with similar semantics.
    
    \item We conduct extensive evaluations and results show that \texttt{Inter}-\texttt{BIN} achieves high accuracy on both basic block level (AUC = 0.99) and function level (precision@1 = 0.85 to 0.96) inputs, significantly outperforms state-of-the-art cross-architecture binary comparison approaches.
    
    \item  We present \emph{CrossMal}, a large-scale IoT malware dataset collected from wild network environments containing 1,878,437 cross-architecture reuse function pairs, involving seven malware families such as \texttt{Mirai}, \texttt{Gafgyt}, and \texttt{Hajime}. Evaluation and case analysis on \emph{CrossMal} prove that \texttt{Inter}-\texttt{BIN} is practical and scalable in real-world IoT scenarios.
\end{itemize}

\section {Problem Defination}
In this section, we formalize the cross-architecture binary similarity comparison problem as follows:

Given two binary code pieces $B_a$ and $B_b$ compiled on different architectures with separate instruction sets, our goal is to compute their semantic similarity score ranging from 0 to 1. 0 represents their semantics are completely different, and 1 denotes they are semantically equivalent. Semantic similarity score can measure the functional similarity of code pieces. If $B_a$ and $B_b$ are semantically equivalent, they will produce exactly the same output when given the same input.

We determine the semantic similarity of $B_a$ and $B_b$ based on their corresponding disassembly instruction sequences $\left(\iota_{1}^{(a)}, \cdots, \iota_{M}^{(a)}\right)$ and $\left(\iota_{1}^{(b)}, \cdots, \iota_{N}^{(b)}\right)$. $M$ and $N$ indicate the length of the two instruction sequences. $\iota_{i}^{(a)}$ denotes the $i$ position assembly instruction of $B_a$. As demonstrated in section \uppercase\expandafter{\romannumeral1}.A, the lexical and syntax expressions of similar functional instruction pairs compiled on different architectures may be completely different, indicating cross-architecture binary similarity comparison is a nontrivial task. 

\section{System Overview}
\texttt{Inter}-\texttt{BIN} is an end-to-end interaction-based cross-architecture binary code similarity comparison system. \figurename \ref {workflow} shows the overall workflow of \texttt{Inter}-\texttt{BIN}, including four modules to implement its functionality:
\begin{itemize}
  \item Multi-feature fusion-based instruction representation module: We disassemble IoT binaries and preprocess the instructions, extract their character sequences and statistics attributes. Then we extract the instruction's spatial features by char-level embedding and a 1-D convolutional layer, and use an opcode embedding layer and an operand feature linear mapping layer to learn deeper semantic information of the instruction. We fuse the learned vectors as the final representation of the instruction.
  
  \item Instruction Sequence Encoding module: This module uses Bi-LSTM to encode instruction sequence streams, generating sequence representations with bidirectional context information. The parameters of instruction representation layers and the sequence encoding layer are shared among binaries of different architectures.
  
  \item Cross-architecture interaction module: This module performs automatic soft alignment of cross-architecture instruction pairs as an inter-sequence interaction schema, which can associate instructions with relevant functional semantics but different lexical and syntax expressions.
  
  \item Binary similarity matching module: We concatenate cross-architecture instruction sequence representations enhanced by the interaction module, and use a fully connected layer to generate binary matching result.
\end{itemize}
We will elaborate on the specific technical implementation of each module in section \uppercase \expandafter{\romannumeral4},  section \uppercase \expandafter{\romannumeral5}, section  \uppercase \expandafter{\romannumeral6}, and section  \uppercase \expandafter{\romannumeral7}.

\section{Multi-feature Fusion-based Instruction Representation}
To avoid the laborious workload and the out-of-vocabulary (OOV) problem of instruction pre-training methods, we design an instruction vectorization approach based on fusing multiple lightweight features, including char-level features, opcode features and operands features. 

The iput of the \emph{multi-feature fusion-based instruction representation module} is 
a pair of disassembled cross-architecture binary snippets $B_a$ and $B_b$ , which can be formalized as $I_a = \left(\iota_{1}^{(a)}, \cdots, \iota_{M}^{(a)}\right)$ and $I_b = \left(\iota_{1}^{(b)}, \cdots, \iota_{N}^{(b)}\right)$. An assembly instruction is composed of an opcode and zero to more operands. The opcode specifies the operation to be conducted, and the operands specify literals, registers, or memory locations of the opcode. For each instruction, we first preprocess it to reduce the lexical gap among instructions of different architectures. We replace string literals, numeric literals, function names, and other symbolic constants with unified identifiers. Then we treat the preprocessed instruction as a whole target for vectorization.

\begin{figure*}[!t]
	\centering
	\includegraphics[width=11cm]{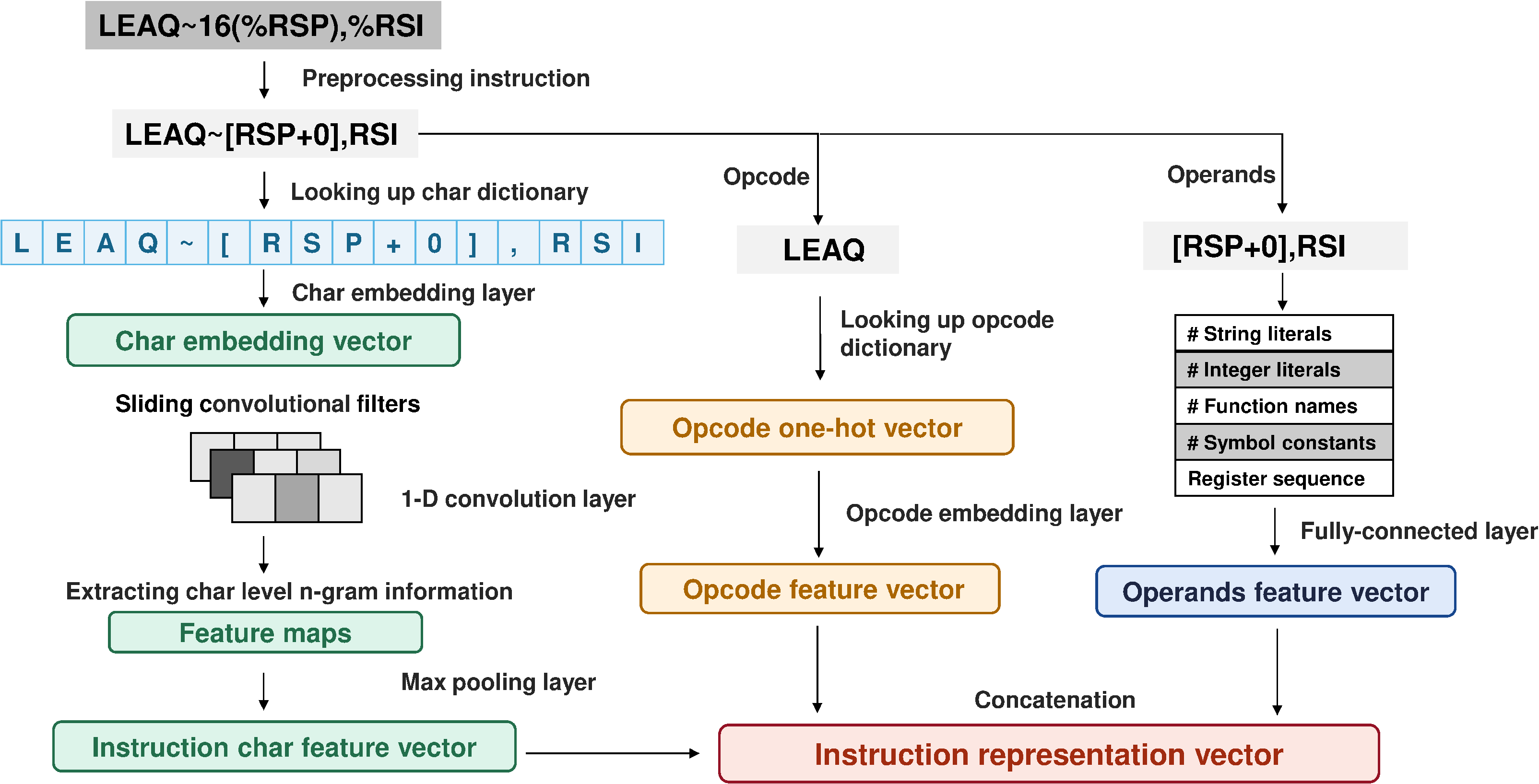}
	\caption{Multi-feature fusion-based instruction representation process.}
	\label{multifeature}
\end{figure*}

\noindent \textbf{Char-level features.}
We create a char dictionary table for the character sequences of the preprocessed assembly instructions. Unlike the large instruction vocabulary used by previous work \cite{zuo2018neural} \cite{massarelli2019SAFE}, the char dictionary table can be maintained at a small scale. The size of our char table is 58, containing 26 English letters, ten digits, and 22 special characters that may appear in the instruction, such as \$, +, -, [, and ].

For the assembly instruction, we look up the position index of each character of the instruction in the char dictionary and express it as a one-hot encoding vector. The characters not appeared in the table are represented as all-zero vectors, while this phenomenon  didn't happen in our evaluation. Then we use a char embedding layer to map the one-hot vectors into discrete dense vectors. The char sequence of an instruction will be translated to a char embedding vector sequence.

We truncate the character vector sequence of the instruction to a fixed length and then use a 1-D convolutional layer to extract its char-level spatial features. The 1-D convolutional layer uses multiple convolution filters to slide on the instruction's character sequence, capturing feature patterns from different perspectives. It is worth mentioning that a convolution kernel of size $n$ can generate a feature map containing char-level n-gram information as Equation 1:
\vspace{-1mm}
\begin{equation}
F_{i}=\sigma\left(\mathbf{w} \cdot \mathbf{c}_{k: k+n-1}+b\right)
\vspace{-1mm}
\end{equation}
$F_i$ is the feature generated by the convolution operation on the characters window $\mathbf{c}_{k: k+n-1}$. $\mathbf{w}$ is the 1-D convolution filter and $b$ is a bias. $\sigma$ is a non-linear activation function. 

Char-level n-gram features are beneficial for the semantic characterization of disassembly instructions. For example, on the x86 architecture, opcodes “$MOV, MOVQ, MOVD$” have similar semantics. When setting $n$ to $3$, their generated char 3-gram features will all contain the “$MOV$” term, which implies that the instructions will execute operations related to data copies. Furthermore, although different CPU architectures have separate instruction sets, some opcodes with similar operations may have similar char-level lexical features. Such as "$ADD, ADDPD, ADDSD$" of the x86 architectures, and "$ADD, ADDS$" of the ARM architecture will perform similar operations. Their char 3-grams will all contain "$ADD$". Our instruction representation layers are shared among different architectures. Similar char n-gram features extracted from disassembly instructions of different architectures will imply their semantic similarity information. It is beneficial for the cross-architecture instruction alignment of the subsequent module.

Completing the convolution operation on each character sliding window of size $n$, the overall character sequence within an instruction of length $l$ will generate a feature map $\mathbf{F}_{char}=\left[F_{1}, F_{2}, \ldots, F_{l-n+1}\right]$.

After the feature pattern extraction by the 1-D convolutional layer, we use a 1-D max-pooling layer to extract the most important char-level feature information $\mathbf{f}_{char}$ in the temporal dimension of the feature map $\mathbf{F}_{char}$: 
\vspace{-1mm}
\begin{equation}
\mathbf{f}_{char}=\max \{\mathbf{F}_{char}\}
\vspace{-1mm}
\end{equation}

\noindent  \textbf{Opcode features.}
To further learn the semantic information of the assembly instructions, we set up the opcode feature extraction layer and the operands feature extraction layer to learn the corresponding representation.

For opcode, we construct an opcode lookup table for each architecture and generate a one-hot representation of the input instruction's opcode type. Then we fed the one-hot vector into an embedding layer to generate the distributed opcode-based feature vector $\mathbf{f}_{opcode}$.

\noindent  \textbf{Operands features.}
For operands, we build a register dictionary table for each architecture, and extract the following statistical attributes:
(1) The number of string literals.
(2) The number of integer literals.
(3) The number of functions names.
(4) The number of other symbolic constants.
(5) The one-hot vectors of the registers sequence.
We concatenate the statistical information of operands and send it to a fully connected layer to generate the operands-based feature vector $\mathbf{f}_{operands}$.

The overall multi-feature fusion-based instruction vectorization process is shown in \figurename \ref {multifeature}. The char-based feature vector $\mathbf{f}_{char}$, opcode-based feature vector $\mathbf{f}_{opcode}$, and the operands-based feature vector $\mathbf{f}_{operands}$ of instruction $i$ are concatenated together as Equation 3 to generate the final output $\mathbf{f}_i$ of the instruction representation module. The vectorized instruction representations will be used as input to the subsequent instruction sequence modeling module.  

\begin{equation}
\mathbf{f}_i = [\mathbf{f}_{char} ; \mathbf{f}_{opcode} ; \mathbf{f}_{operands}]
\end{equation}
The output of the \emph{multi-feature fusion-based instruction representation module} is the instruction representation sequences $\mathbf{F}_{a} = \left(\mathbf{f}_{1}^{(a)}, \cdots, \mathbf{f}_{M}^{(a)}\right)$ and $\mathbf{F}_{b} = \left(\mathbf{f}_{1}^{(b)}, \cdots, \mathbf{f}_{N}^{(b)}\right)$.

\section{Instruction Sequence Encoding}
After the \emph{multi-feature fusion-based instruction representation module}, the binary pieces pair $B_a$ and $B_b$ are converted into $\mathbf{F}_{a} = \left(\mathbf{f}_{1}^{(a)}, \cdots, \mathbf{f}_{M}^{(a)}\right)$ and $\mathbf{F}_{b} = \left(\mathbf{f}_{1}^{(b)}, \cdots, \mathbf{f}_{N}^{(b)}\right)$ as the input of the \emph{instruction sequence encoding module}. $\mathbf{f}_{i}^{(a)}$ denotes the representation of the $i$ position instruction of $B_a$. We use Bi-LSTM to encode the two instruction sequences. The LSTM network can alleviate the gradient disappearance and gradient explosion problems of the basic RNN recursive architecture and model long-distance dependence with better performance \cite{hochreiter1997long}. To better model the context-dependent information in the instruction sequence, we consider Bi-LSTM, which simultaneously maintains a forward hidden state $\overrightarrow{\mathbf{h}_{t}}$ and a backward hidden state $\overleftarrow{\mathbf{h}_{t}}$ at time step $t$. The two hidden layer states are used to model the preceding and following information of current instruction, respectively. The overall $t$-th step update process of Bi-LSTM hidden layer representation can be formalized as Equation 4:
\vspace{-1mm}
\begin{equation}
\stackrel{\leftrightarrow}{\mathbf{h}}_{t}=\operatorname{Bi-LSTM}\left(\mathbf{f}_{t}, \stackrel{\leftrightarrow}{\mathbf{h}}_{t-1}\right)
\vspace{-1mm}
\end{equation}
$\mathbf{f}_{t}$ represents the vector of the $t$-th instruction in the sequence, and $\stackrel{\leftrightarrow}{\mathbf{h}}_{t}$ represents the Bi-LSTM hidden state of the $t$-th step, which is formed by concatenating the forward and backward hidden states. We concatenate the bidirectional representations of each Bi-LSTM hidden state as the encoded instruction sequence embeddings of binary snippets $B_a$ and $B_b$, represented as $\mathbf{H}_{a} = \left(\mathbf{h}_{1}^{(a)}, \cdots, \mathbf{h}_{M}^{(a)}\right)$ and $\mathbf{H}_{b} = \left(\mathbf{h}_{1}^{(b)}, \cdots, \mathbf{h}_{N}^{(b)}\right)$. $\mathbf{H}_{a}$ and $\mathbf{H}_{b}$ are the output of the \emph{instruction sequence encoding module}.

\section{Cross-architecture Interaction}
\begin{figure*}[!t]
	\centering
	\includegraphics[width=11.5cm]{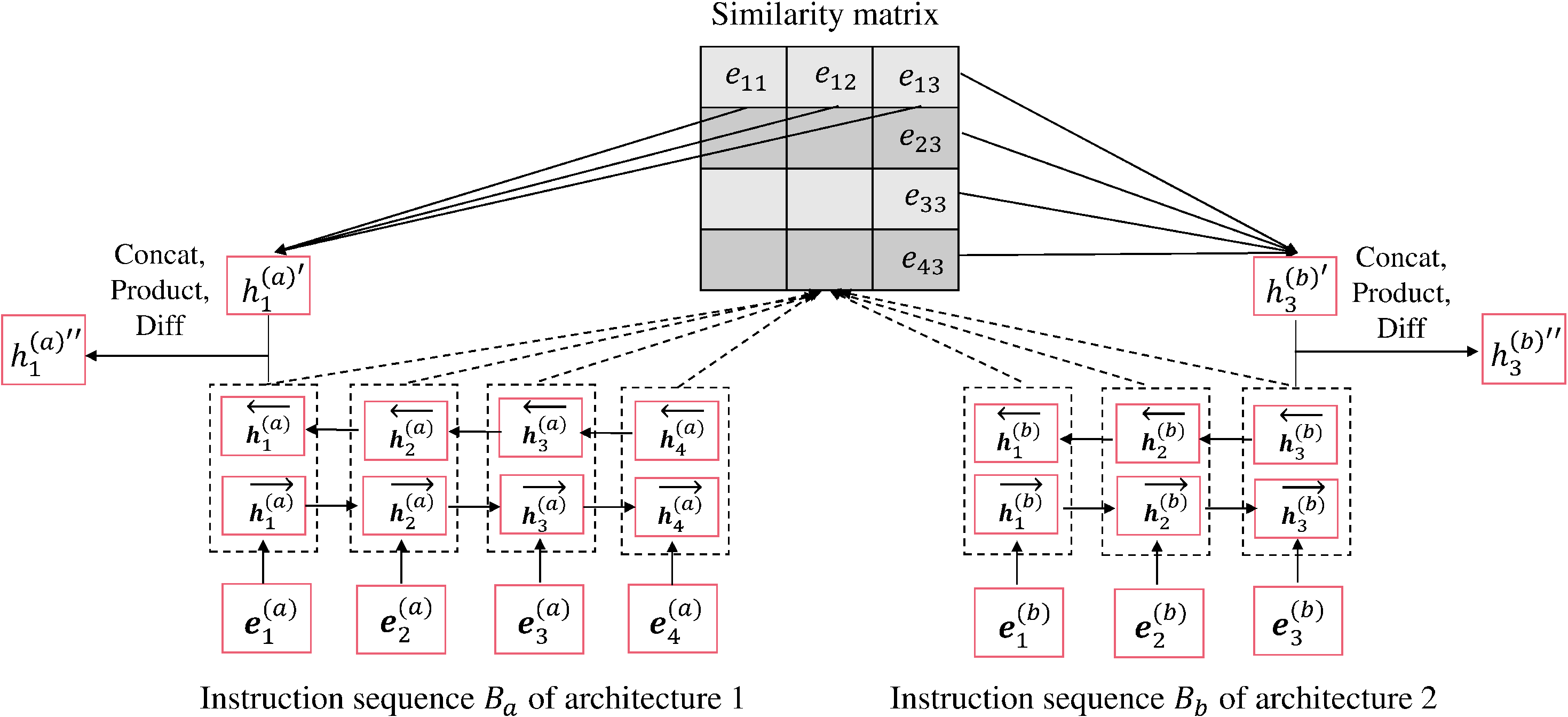}
	\caption{Co-attention based soft alignment of instruction pairs.}
	\label{coattn}
\end{figure*}

In this section, we introduce the semantic interaction between cross-architecture instruction sequence representations. The input of our \emph{cross-architecture interaction module} is the instruction vector sequences $\mathbf{H}_{a} = \left(\mathbf{h}_{1}^{(a)}, \cdots, \mathbf{h}_{M}^{(a)}\right)$ and $\mathbf{H}_{b} = \left(\mathbf{h}_{1}^{(b)}, \cdots, \mathbf{h}_{N}^{(b)}\right)$ generated by the \emph{instruction sequence encoding module}. Our goal is to flexibly and accurately associate instruction pairs with similar functions but different syntax expressions. We use the co-attention mechanism to achieve inter-sequence interactions, which can realize automatic soft alignment between instruction pairs by modeling their semantic correlation.

\figurename \ref {coattn} shows the calculation of our co-attention based instruction sequence interaction process. For the instruction sequences pair $B_a$ and $B_b$ from different architectures, we first calculate the instruction semantical similarity matrix by a bilinear layer as Equation 5:
\vspace{-1mm}
\begin{equation}
e_{i j}=\mathbf{h}_{i}^{(a)} \mathbf{W}_{s} \mathbf{h}_{j}^{(b)}
\vspace{-1mm}
\end{equation}
\vspace{-1mm}\\Then we get the attentive representation of $B_b$ under the guidance of $B_a$, and vice versa:
\begin{equation}
{\mathbf{h}_{i}^{(a)}}^{\prime}=\sum_{j=1}^{n} \frac{\exp \left(e_{i j}\right)}{\sum_{k=1}^{n} \exp \left(e_{i k}\right)} \mathbf{h}_{j}^{(b)}
\vspace{-1mm}
\end{equation}
\vspace{-1mm}
\begin{equation}
{\mathbf{h}_{j}^{(b)}}^{\prime}=\sum_{i=1}^{m} \frac{\exp \left(e_{i j}\right)}{\sum_{k=1}^{m} \exp \left(e_{k j}\right)} \mathbf{h}_{i}^{(a)}
\vspace{-1mm}
\end{equation}
$\mathbf{h}_{i}^{(a)}$ and $\mathbf{h}_{j}^{(b)}$ indicate the vectors of the instruction $\iota_{i}^{(a)}$ and $\iota_{j}^{(b)}$ in $B_a$ and $B_b$ generated by the sequence encoding module. $e_{i j}$ represents their semantic similarity. $\mathbf{W}_{s}$ is the parameter matrix of the bilinear layer. For the instruction representation $\mathbf{h}_{i}^{(a)}$ in $B_a$, we use the semantic similarity between it and each instruction vector $\mathbf{h}_{j}^{(b)}$ of $B_b$ to calculate the attention coefficient through softmax. Then we perform a weighted summation to obtain the attentive representation ${\mathbf{h}_{i}^{(a)}}^{\prime}$. The instruction representation $\mathbf{h}_{j}^{(b)}$ of $B_b$ is updated in the same way. The cross-architecture sequence interaction process is performed symmetrically in parallel.

To further enhance the effect of the interaction module, we follow \cite{conneau2017supervised} to perform multiple combination ways of the original instruction representation $\mathbf{h}_{i}^{(a)}$ and the attentive representation ${\mathbf{h}_{i}^{(a)}}^{\prime}$, including concatenation, element-wise difference, and element-wise product. We concatenate the three enhanced results as ${{\mathbf{h}_{i}}^{(a)}}^{\prime \prime}$ and ${{\mathbf{h}_{j}}^{(b)}}^{\prime \prime}$, then generate the final instruction representations $\tilde{{\mathbf{h}_{i}}}^{(a)}$ and $\tilde{{\mathbf{h}_{j}}}^{(b)}$ by a one-layer feed-forward neural network $F$ as Equations 8-11: 
\vspace{-1mm}
\begin{equation}
{{\mathbf{h}_{i}}^{(a)}}^{\prime \prime}=\left[{\mathbf{h}_{i}}^{(a)} ; {{\mathbf{h}_{i}}^{(a)}}^{\prime} ; {\mathbf{h}_{i}}^{(a)}-{{\mathbf{h}_{i}}^{(a)}}^{\prime} ; {\mathbf{h}_{i}}^{(a)} \odot {{\mathbf{h}_{i}}^{(a)}}^{\prime}\right]
\vspace{-1mm}
\end{equation}
\vspace{-1mm}
\begin{equation}
\tilde{{\mathbf{h}_{i}}}^{(a)}=F\left({{\mathbf{h}_{i}}^{(a)}}^{\prime \prime}\right) 
\vspace{-1mm}
\end{equation}
\vspace{-1mm}
\begin{equation}
{{\mathbf{h}_{j}}^{(b)}}^{\prime \prime}=\left[{\mathbf{h}_{j}}^{(b)} ; {{\mathbf{h}_{j}}^{(b)}}^{\prime} ; {\mathbf{h}_{j}}^{(b)}-{{\mathbf{h}_{j}}^{(b)}}^{\prime} ; {\mathbf{h}_{j}}^{(b)} \odot {{\mathbf{h}_{j}}^{(b)}}^{\prime}\right]
\vspace{-1mm}
\end{equation}
\vspace{-1mm}
\begin{equation}
\tilde{{\mathbf{h}_{j}}}^{(b)}=F\left({{\mathbf{h}_{j}}^{(b)}}^{\prime \prime}\right)
\vspace{-1mm}
\end{equation}
The enhanced instruction vector sequences  $\mathbf{H_a}^{\prime \prime} = \left({{\mathbf{h}_{1}}^{(a)}}^{\prime \prime}, \cdots, {{\mathbf{h}_{M}}^{(a)}}^{\prime \prime}\right)$ and $\mathbf{H_b}^{\prime \prime} = \left({{\mathbf{h}_{1}}^{(b)}}^{\prime \prime}, \cdots, {{\mathbf{h}_{N}}^{(b)}}^{\prime \prime}\right)$ act as the output of the \emph{cross-architecture interaction module}.

\section{Binary Similarity Matching}
The input of the \emph{binary similarity matching module} is the instruction vector sequences $\mathbf{H_a}^{\prime \prime}$ and $\mathbf{H_b}^{\prime \prime}$ generated by the \emph{cross-architecture interaction module}. We use the summation function to aggregate the instruction vectors of each sequence enhanced by the cross-architecture interaction module, as shown in Equations 12–13. ${\mathbf{r}}^{(a)}$ and ${\mathbf{r}}^{(b)}$ represent the final representation of binary snippets $B_a$ and $B_b$. We also tried other instruction sequence aggregation methods but observed no further improvement. The ultimate vectors contain bidirectional context information and the semantic interaction information of cross-architecture instructions.
\vspace{-1mm}
\begin{equation}
{\mathbf{r}}^{(a)}=\sum_{i=1}^{M} \tilde{{\mathbf{h}_{i}}}^{(a)}
\vspace{-1mm}
\end{equation}
\vspace{-1mm}
\begin{equation}
{\mathbf{r}}^{(b)}=\sum_{j=1}^{N} \tilde{{\mathbf{h}_{j}}}^{(b)}
\vspace{-1mm}
\end{equation}

We concatenate the final sequence representations and feed them into fully connected layers. Then we deploy a \emph{softmax} output layer to generate the binary similarity comparison result. We employ the cross-entropy function to calculate loss $L$ as Equation 14. $P$ denotes the total number of binary pairs, and $c$ denotes the category of the matching result, 1 for similar and 0 for dissimilar. $y_{p, c}$ is the model’s prediction of the probability that the similarity comparison result of the binary pair $p$ is $c$, and $\hat{y}_{p, c}$ denotes the ground-truth.
\begin{equation}
L=-\sum_{p=1}^{P} \sum_{c=0,1} \hat{y}_{p, c} \log \left(y_{p, c}\right)
\end{equation}

\section{Experimental Evaluation}
We implement our solution as a universal cross-architecture binary similarity comparison prototype, \texttt{Inter}-\texttt{BIN}. In this section, we evaluate \texttt{Inter}-\texttt{BIN} on two input granularities: basic block level and function level. First, we describe the datasets and evaluation metrics used in our experiments (section \uppercase\expandafter{\romannumeral8}.A). Next, we compare our multi-feature fusion-based instruction representation module with the instruction pre-training approach (section \uppercase\expandafter{\romannumeral8}.B). Then we perform ablation studies to evaluate how the designed core components contribute to \texttt{Inter}-\texttt{BIN}'s performance
improvements (section \uppercase\expandafter{\romannumeral8}.C), and we adjust the hyper-parameters to achieve the optimal performance (section \uppercase\expandafter{\romannumeral8}.D). We evaluate \texttt{Inter}-\texttt{BIN}'s multi-granularity performance and make comparisons with state-of-the-art binary similarity comparison approaches, and carry out specific studies on the improved cases (section \uppercase\expandafter{\romannumeral8}.E and section \uppercase\expandafter{\romannumeral8}.F). Furthermore, we evaluate the scalability and efficiency of \texttt{Inter}-\texttt{BIN} on our collected real-world cross-architecture IoT malware reuse function matching dataset \emph{CrossMal} (section \uppercase\expandafter{\romannumeral8}.G).

We implement our prototype using the \texttt{PyTorch}-\texttt{matchzoo} framework \cite{Guo2019matchzoo}.

\subsection{Dataset and Metrics}
We collect three datasets in our experiments: Dataset1 is a collection of basic blocks for evaluating \texttt{Inter}-\texttt{BIN}'s characterization and comparison ability of small and common binary code pieces. Dataset2 is function level for evaluating \texttt{Inter}-\texttt{BIN} on binary snippets with richer semantics and more complex structure. Dataset3 is used to evaluate the feasibility of \texttt{Inter}-\texttt{BIN} in real-world IoT scenarios.

\begin{itemize}
	\item \textbf{Dataset1:} We use the dataset provided by the state-of-the-art approach \texttt{INNEREYE} \footnote{\url{https://nmt4binaries.github.io/}} to evaluate \texttt{Inter}-\texttt{BIN} at the basic block level. The dataset compiles \texttt{OpenSSL} and four popular Linux packages, including \texttt{coreutils}, \texttt{findutils}, \texttt{diffutils}, and \texttt{binutils}, on x86 and ARM platforms. It contains 56,082 similar basic block pairs and 55,937 dissimilar pairs.
	
	\item \textbf{Dataset2:} We expand Dataset1 into a function level binary similarity comparison dataset, using two compilers, \texttt{clang} and \texttt{GCC}, and four optimization levels from O0 to O3. Dataset2 contains a total of 485,025 function pairs. 
	
	\item \textbf{Dataset3:} We collect malware targeting IoT devices (routers and video surveillance devices) by the \texttt{IoTCMal} honeypot \cite{wang2020iotcmal} to evaluate \texttt{Inter}-\texttt{BIN} in real-world scenarios. The captured instances involve seven malware families, including \texttt{Mirai}, \texttt{Hajime}, \texttt{Gafgyt}, \texttt{XXorDDoS}, \texttt{Dofloo}, \texttt{Ddostf}, \texttt{Mining}, and spread on different CPU architectures. We select malware from x86, ARM, and MIPS and annotate the malware pairs compiled from the same source code based on their code structure and external information, including binary names and the captured IP address. We disassemble binaries by radare2 \footnote{\url{https://www.radare.org/}} and extract function level instruction sequences to construct a cross-architecture binary comparison dataset containing 1,878,437 function pairs, and we name it as \emph{CrossMal}.	
\end{itemize}

\noindent \textbf{Ground truth.}
The ground truth of Dataset1 is labelled by \texttt{INNEREYE}. They modified the LLVM-backend to annotate the boundaries of basic blocks and assign a unique ID for each assembly block. The same ID indicates that the assembly blocks are compiled from the same piece of source code. 

For Dataset2 and Dataset3, the function level binary comparison datasets, we use the binary name and function name as the unique ID to identify functions compiled from the same source code. For each query function, we randomly sample the function with the same ID from the target architectures as the positive candidate function, and then sample $num\_neg$ functions with different IDs as the negative candidates.

\noindent \textbf{Evaluation metrics.}
We use the following metrics in evaluation: For basic block pairs, we set up a classification task evaluated by accuracy and Area Under the Curve (AUC) metrics, which is commonly used in previous basic block comparison works \cite{zuo2018neural} \cite{redmond2018cross}. For function granularity evaluation, we set up a ranking task to meet the needs of real-world scenarios such as malware reuse modules detection and vulnerability function discovery. We treat each assembly function as a query, perform a one-to-many comparison with the target functions, and rank the similarity comparison results. We set one positive pair for each query and use precision@1 and mean reciprocal rank (MRR) metrics for evaluation. precision@1 calculates the correct rate of function matching result ranked at position 1. MRR is defined as Equation 15:
\begin{equation}
M R R=\frac{1}{|F|} \sum_{f \in F} \frac{1}{\operatorname{rank(f)}}
\end{equation}
$|F|$ is the number of query functions, $\operatorname{rank(f)}$ is the position of the first correctly matched function of query $f$.

\subsection{Comparison with Instruction Pre-training Approach}
In this section, we compare \texttt{Inter}-\texttt{BIN}'s multi-feature fusion-based instruction representation module with the instruction pre-training approach and char feature-based instruction encoding methods.

we use the released instruction embedding files of dimensions (dims) 50, 100, and 150 provided by \texttt{INNEREYE} \footnote{\url{https://nmt4binaries.github.io/}\label{innereye_web}}, which were trained on a large-scale external code corpus by the skip-gram model. For char feature-based instruction encoding, we test the appearance frequencies of the chars and the combination of their frequency and position information, including the position of the char's first and last appearance. We also evaluate char-level 2-gram and 3-gram features. Table \ref {char_cmp} shows the comparison results on Dataset1. 

\begin{table}
	\caption{Performance of different instruction representation approaches.}
	\label{char_cmp}
	\centering
	\renewcommand{\arraystretch}{1.1}
	\begin{tabular}{|m{110pt}<{\centering}|m{40pt}<{\centering}|m{45pt}<{\centering}|}
		\hline
		Approaches & Accuracy\\
		\hline
		Pre-trained embedding-50 dims & 0.9585 \\
		Pre-trained embedding-100 dims & 0.9633 \\
		Pre-trained embedding-150 dims & 0.9628 \\
		
		\hline
		Char frequency & 0.9370 \\
		Char frequency with position & 0.9515 \\
		Char 2-gram & 0.9260 \\
		Char 3-gram & 0.9103 \\
		
		\hline
		\texttt{Inter}-\texttt{BIN} & \textbf{0.9757} \\
		\hline
	\end{tabular}
\end{table}

From the table, \texttt{Inter}-\texttt{BIN}'s instruction representation module significantly outperforms the hard-encoded char features. Compared with \texttt{INNEREYE}'s heavyweight instruction pre-training, the char spatial features and instruction semantic features extracted by \texttt{Inter}-\texttt{BIN} can achieve better results, and the implementation way is very efficient.

\subsection{Ablation Studies}
In this section, we design ablation studies to evaluate the effects of \texttt{Inter}-\texttt{BIN}’s core components.

\begin{table}
	\caption{Ablation Studies of \texttt{Inter}-\texttt{BIN}.}
	\label{add_ablation_table}
	\centering
	\renewcommand{\arraystretch}{1.1}
	\begin{tabular}{|m{110pt}<{\centering}|m{40pt}<{\centering}|m{45pt}<{\centering}|}
		\hline
		Setting & Dataset1 Accuracy & Dataset2 Precision@1\\
		\hline
		(- Char-based features) & 0.9662 & 0.9050 \\
		(- Opcode-based features) & 0.9669 & 0.9054 \\
		(- Operands-based features) & 0.9706 & 0.9672 \\
		(- Backward LSTM layer) & 0.9605 & 0.9575 \\
		(- Co-attention mechanism) & 0.9428 & 0.8951 \\
		
		\hline
		\texttt{Inter}-\texttt{BIN} & \textbf{0.9757} & \textbf{0.9691} \\
		\hline
	\end{tabular}
\end{table}

We first study the effectiveness of different instruction features in the \emph{multi-feature fusion-based instruction representation module}. Table \ref {add_ablation_table} shows the performance of \texttt{Inter}-\texttt{BIN} when removing the char-based features, opcode features, and operands features, respectively. It can be seen that Dataset1 is not very sensitive to the instruction feature ablation, and we can achieve acceptable basic block comparison accuracy on each setting. On the Dataset2 O0 subset, removing char-based spatial features or the opcode embedding features significantly impacts the function matching result. The precision@1 value reduces 6.48 and 6.37 points, respectively, and removing the operands features results in a slight performance decrease.

Next, in our \emph{instruction sequence encoding module}, we use a Bi-LSTM encoder to extract the forward and backward context information of the embedded instruction sequence. When removing the backward LSTM layer,  the binary similarity comparison performance on Dataset1 and Dataset2 O0 subset decreased by 1.52 and 1.16 points in accuracy and precision@1, respectively. It shows that adding backward sequential information positively impacts the semantic modeling of disassembly code snippets.

Finally, we evaluate the influence of the co-attention based instruction alignment in the \emph{cross-architecture interaction module}. As shown in table \ref {add_ablation_table}, when the instruction sequence interaction process is removed, the performance of \texttt{Inter}-\texttt{BIN} on Dataset1 and Dataset2 O0 subset has a significant drop, the accuracy and precision@1 value are reduced by 3.29 and 7.40 points, respectively. This proves that the cross-architecture instruction alignment mechanism is necessary for the performance improvement of binary similarity comparison.

\subsection{Parameter Selection}
We adjust the following vital hyper-parameters of \texttt{Inter}-\texttt{BIN} to achieve the optimal performance: a) The RNN architecture  $\mathbb{R}$ of the instruction sequence encoding module. b) The hidden dimensions $\mathbb{H}$ of RNN layers. c) The attention function $\mathbb{A}$ of the cross-architecture interaction module. d) The enhancement mode $\mathbb{E}$ of the interaction module. We discuss how to choose appropriate values for these hyper-parameters.

a) The RNN architecture $\mathbb{R}$ of the instruction sequence encoding module: We evaluate different RNN variants for instruction sequence encoding, including LSTM, Bi-LSTM, Bi-GRU, and multi-layer Bi-LSTM.

b) The hidden dimensions $\mathbb{H}$ of RNN layers: $\mathbb{H}$ defines the hidden state dimension generated by each RNN layer. We vary $\mathbb{H}$ in the range of $\left\{16,32,64,128,256,512 \right\}$.

c) The attention function $\mathbb{A}$ of the cross-architecture interaction module: We evaluate dot-product attention, scaled dot-product attention, cosine attention, and bilinear attention.

d) The enhancement mode $\mathbb{E}$ of the interaction module: $\mathbb{E}$ defines the enhancement way of the inter-sequence interaction results to the original sequence representations. We test concatenation, element-wise product, element-wise difference, and their combinations.

\begin{figure*}[t]
	\centering
	\quad
	\subfigure[RNN architecture type ($\mathbb{R}$)]{
		\centering
		\includegraphics[width=4.5cm]{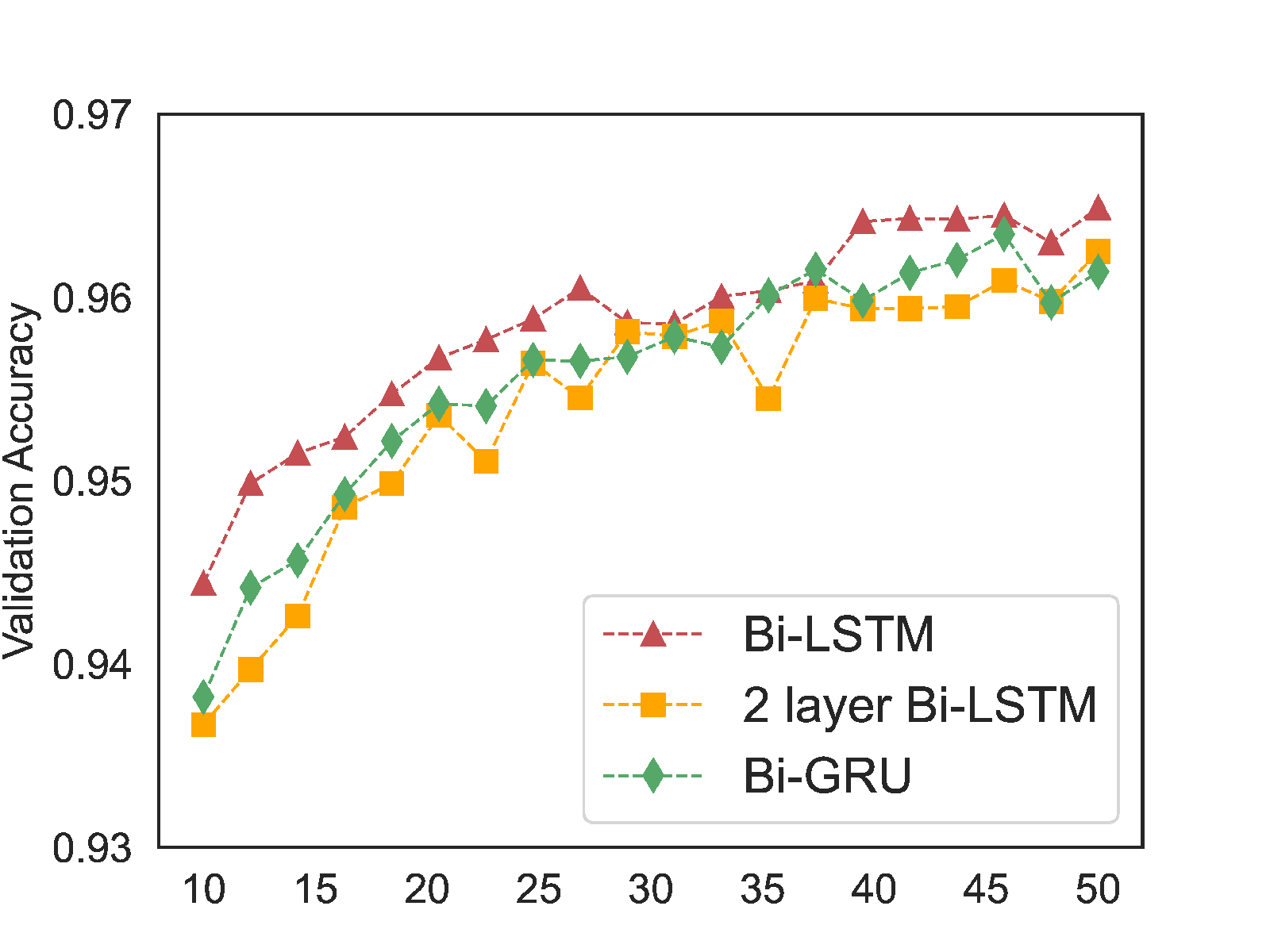}
	}
	\hspace{-10mm}
	\quad
	\subfigure[RNN hidden dimensions ($\mathbb{H}$)]{
		\centering
		\includegraphics[width=4.5cm]{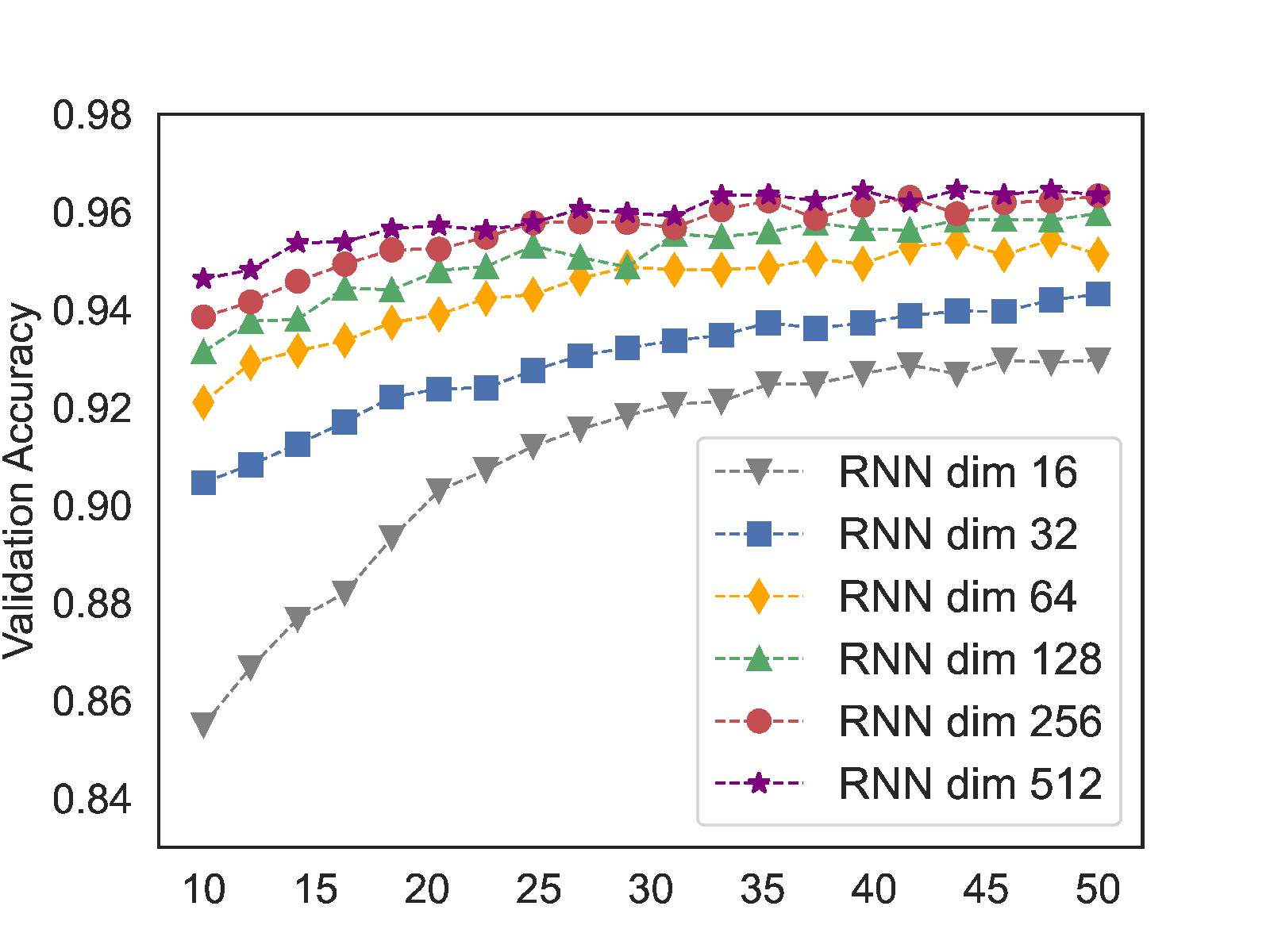}
	}
	\hspace{-10mm}
	\quad
	\subfigure[Attention function ($\mathbb{A}$)]{
		\centering
		\includegraphics[width=4.5cm]{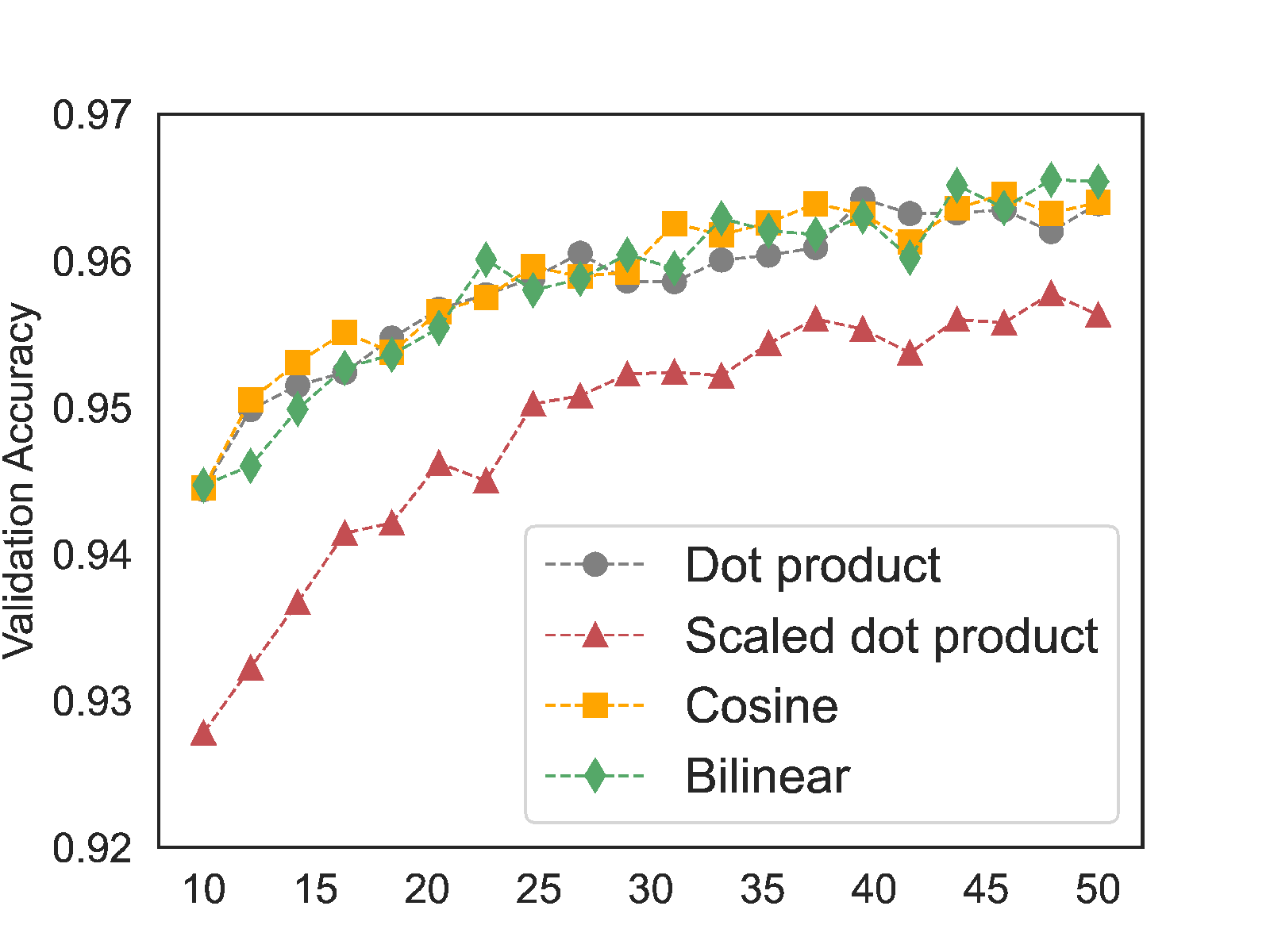}
	}
	\hspace{-10mm}
	\quad
	\subfigure[Interaction enhancement ($\mathbb{E}$)]{
		\centering
		\includegraphics[width=4.5cm]{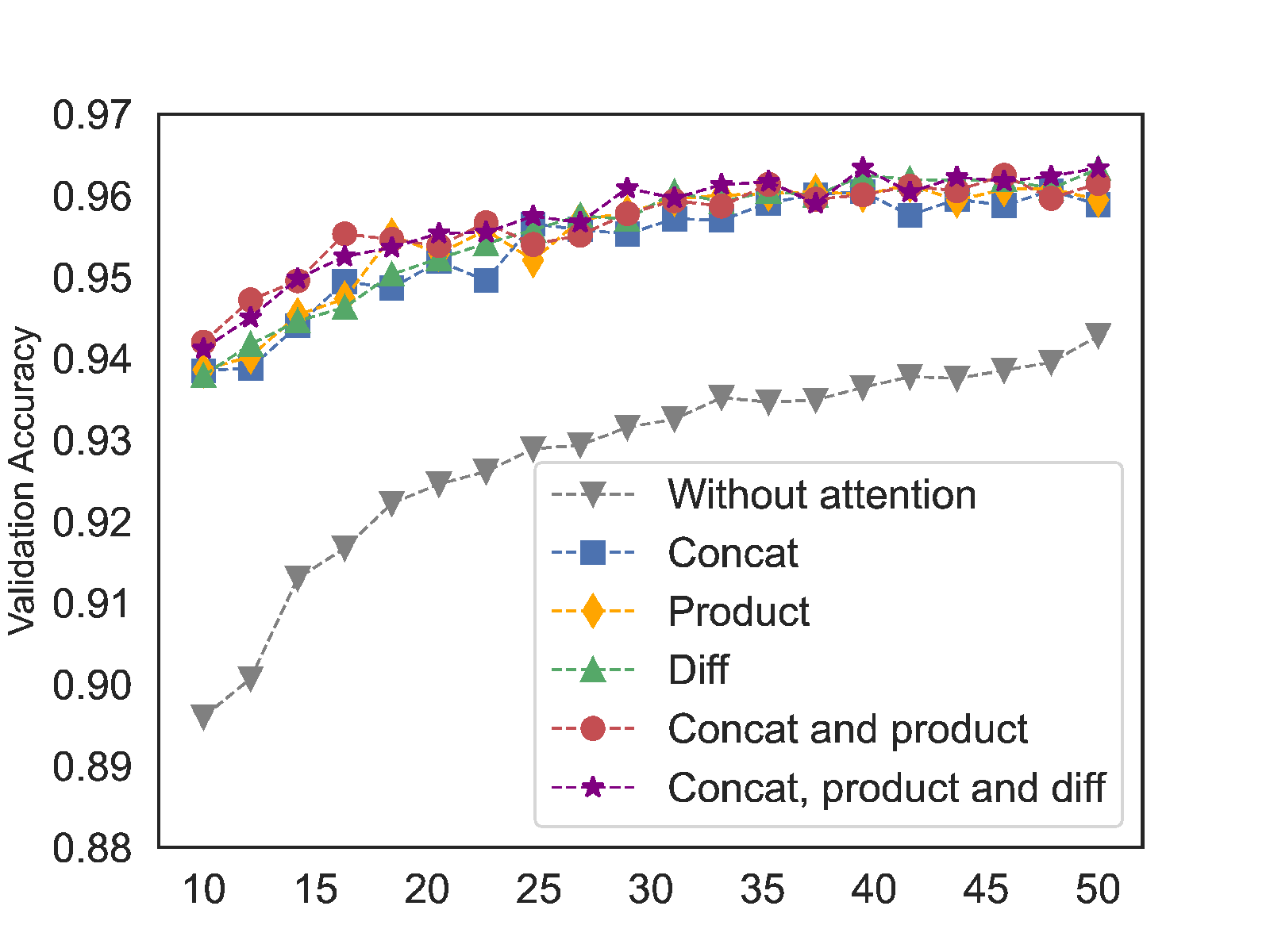}
	}
	\caption{Parameter selection for \texttt{Inter}-\texttt{BIN} on Dataset1}
	\label{params_dt1}
	\vspace*{-1.2\baselineskip}
\end{figure*}

\begin{figure*}[t]
	\centering
	\quad
	\subfigure[RNN architecture type ($\mathbb{R}$)]{
		\includegraphics[width=4.5cm]{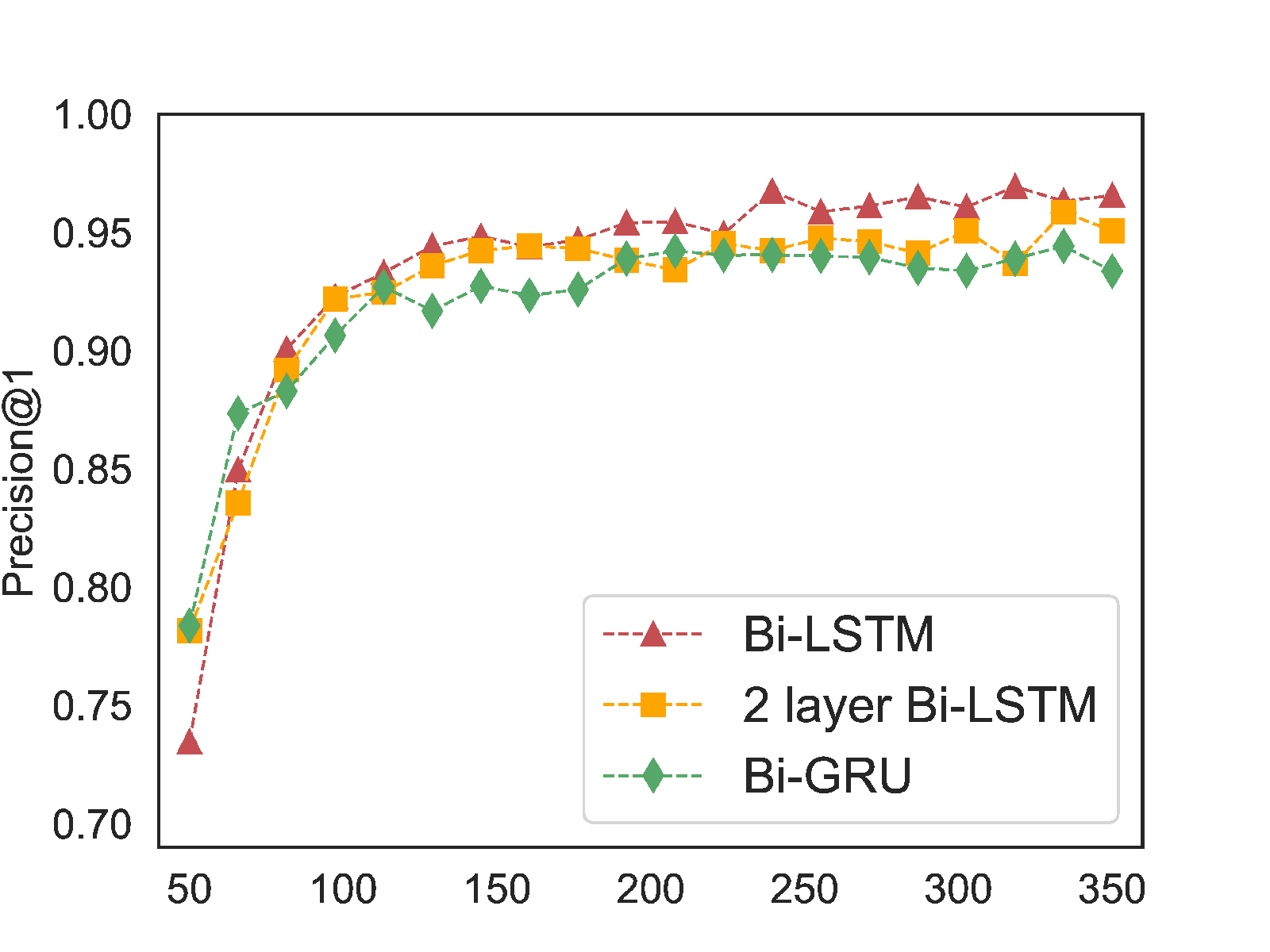}
	}
	\hspace{-10mm}
	\quad
	\subfigure[RNN hidden dimensions ($\mathbb{H}$)]{
		\includegraphics[width=4.5cm]{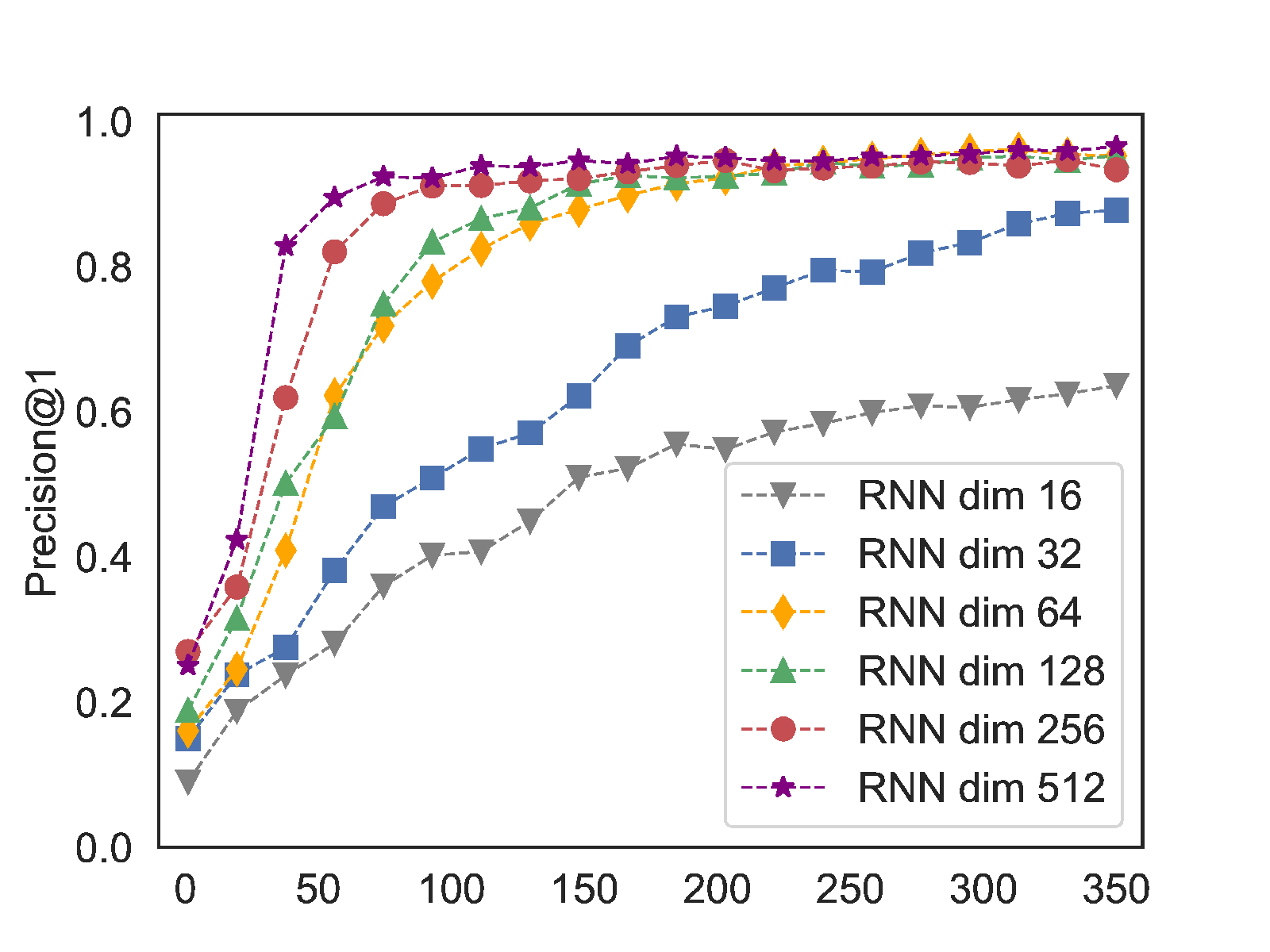}
	}
	\hspace{-10mm}
	\quad
	\subfigure[Attention function ($\mathbb{A}$)]{
		\includegraphics[width=4.5cm]{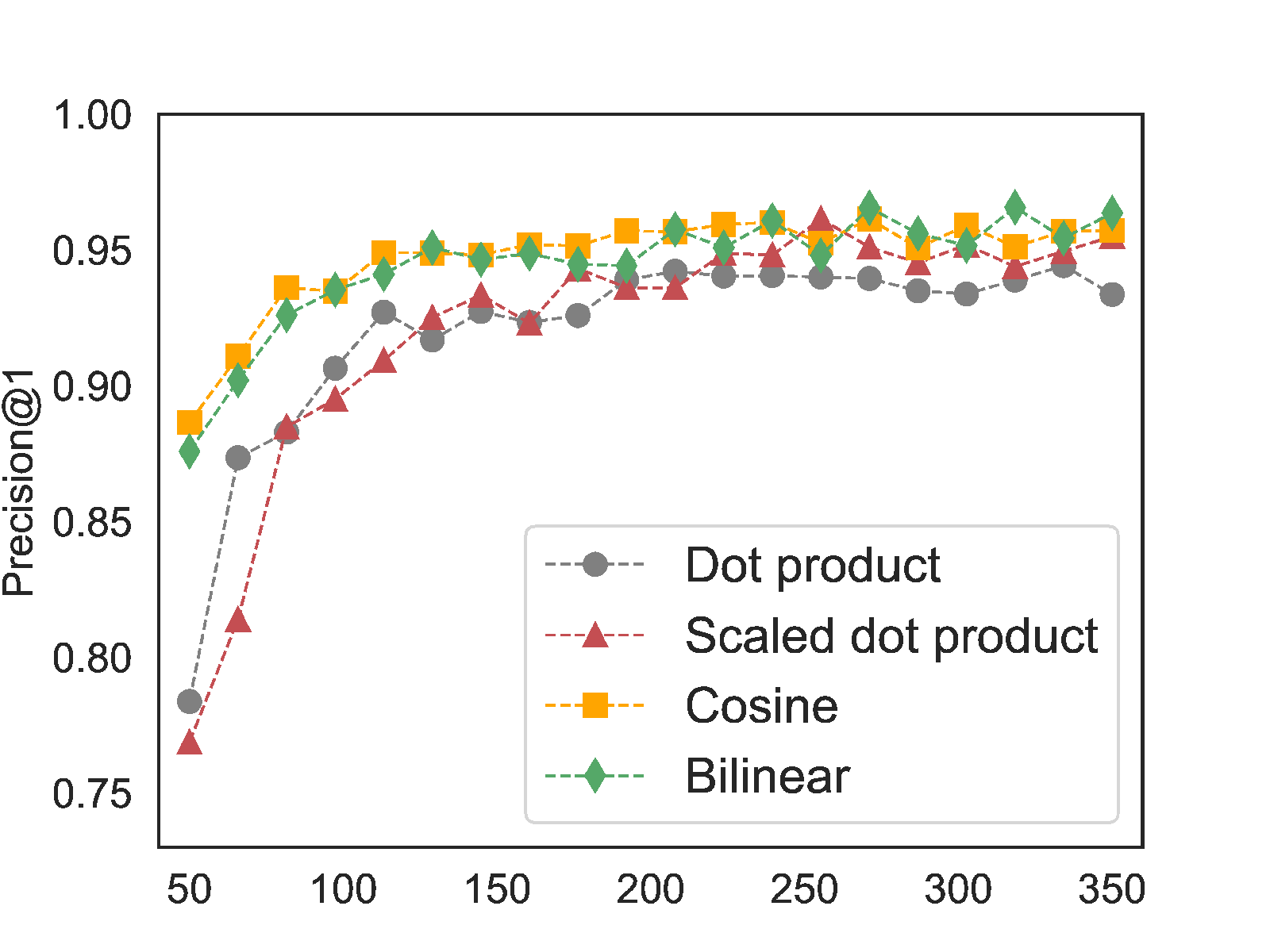}
	}
	\hspace{-10mm}
	\quad
	\subfigure[Interaction enhancement ($\mathbb{E}$)]{
		\includegraphics[width=4.5cm]{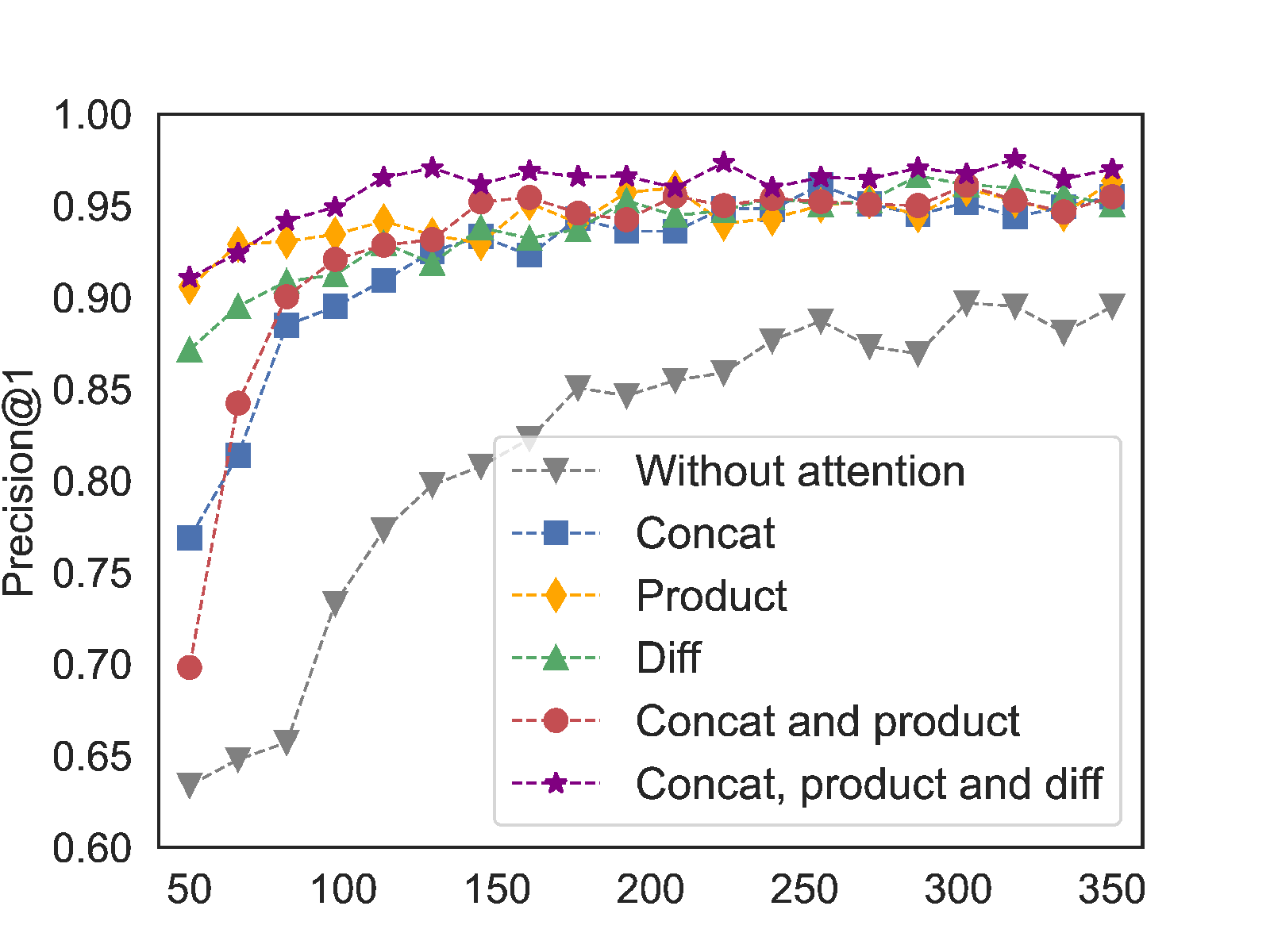}
	}
	\caption{Parameter selection for \texttt{Inter}-\texttt{BIN} on Dataset2}
	\label{params_dt2}
	\vspace*{-1.2\baselineskip}
\end{figure*}

\figurename \ref{params_dt1} and \figurename \ref{params_dt2} show \texttt{Inter}-\texttt{BIN}'s validation accuracy on Dataset1 and precision@1 on Dataset2 O0 validation subset under different hyper-parameters settings. The abscissa indicates the number of training epochs. \figurename \ref{params_dt1}. (a) and \figurename \ref{params_dt2}. (a) show that Bi-LSTM performs slightly better than other RNN variants, and more Bi-LSTM layers doesn't show performance improvement. From \figurename \ref{params_dt1}. (b) and \figurename \ref{params_dt2}. (b), a larger Bi-LSTM hidden dimension has a stronger expressive ability. For the trade-off of accuracy and efficiency, we finally set $\mathbb{H}$ to 256. From \figurename \ref{params_dt1}. (c) and \figurename \ref{params_dt2}. (c), \emph{Bilinear} attention has a slight advantage in network fitting speed and similarity matching results. From \figurename \ref{params_dt1}. (d) and \figurename \ref{params_dt2}. (d), the combinations of three enhancement way can achieve better results.

Other implementation details of \texttt{Inter}-\texttt{BIN} are as follows: The parameters of our neural network are optimized by Adam with a learning rate of 1e-4. The kernel size of the 1-D convolutional layer is set to 3, and the number of filters is set to 64. The hidden dimension of the opcode embedding layer is set to 8 for the basic block level and 64 for the function level, and the dimension of the operands mapping layer is set to 8. The MLP classifier of the binary matching module contains two fully connected layers. Their hidden dimensions are set to 512 and 256, respectively. On our validation sets, the performance of \texttt{Inter}-\texttt{BIN} is not sensitive to these hyper-parameters, and the above settings can achieve the best results.

\subsection{Basic block level Evaluation}
In this section, we evaluate \texttt{Inter}-\texttt{BIN} at the basic block level and compare it with state-of-the-art approaches. Then we perform visualization analysis of our cross-architecture interaction module to show the performance of the automatic instruction soft alignment mechanism. Finally, we analyze the false-positive and false-negative cases generated by the previous approaches but can be avoided by \texttt{Inter}-\texttt{BIN}.

\subsubsection{Comparison with State-of-the-Arts}
We evaluate \texttt{Inter}-\texttt{BIN} on Dataset1 and compare it with state-of-the-art cross-architecture binary matching approaches \texttt{Gemini} \citep{xu2017neural}, \texttt{INNEREYE} \cite{zuo2018neural}, and the work of Redmon et al. \cite{redmond2018cross}. These methods are all related to cross-architecture basic block characterization and comparison. Their implementations are as follows:
\begin{itemize}
	\item \texttt{Gemini} \citep{xu2017neural} uses syntax features like string constants, numeric constants, and the number of instructions to represent a basic block. We use the SVM classifier with \texttt{RBF} kernel for \texttt{Gemini}'s basic blocks comparison \cite{zuo2018neural}.
	
	\item \texttt{INNEREYE} \cite{zuo2018neural} separately trains two LSTMs to encode instruction streams and then uses distance metric to measure the similarity of cross-architecture binary snippets.
	
    \item Redmon et al. \cite{redmond2018cross} perform hard alignment of cross-architecture instructions based on their position indexes, then use a joint learning approach with mono-architecture and cross-architecture objectives to learn instruction embeddings.
\end{itemize}

\begin{figure}[!t]
	\centering
	\includegraphics[width=6cm]{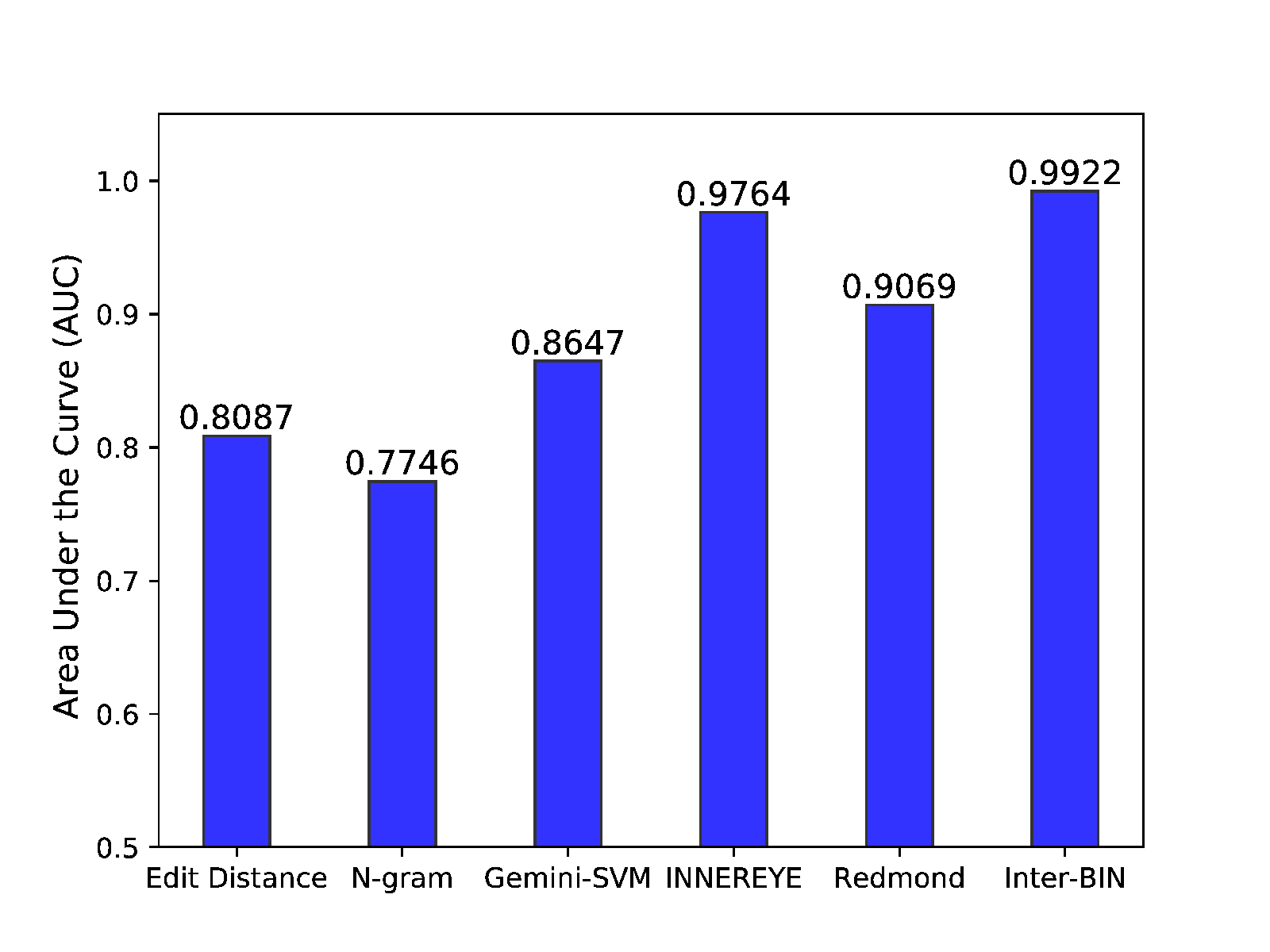}
	\caption{Basic block level cross-architecture binary similarity comparison results on Dataset1.}
	\label{bbcmp}
\end{figure}

We use the instruction embedding files of \texttt{INNEREYE} \textsuperscript{\ref{innereye_web}} and the work of Redmond et al. \footnote{\url{https://github.com/nlp-code-analysis/cross-arch-instr-model}} for replication. The RNN hidden states of \texttt{Inter}-\texttt{BIN} and \texttt{INNEREYE} are all set to 256 dimensions, the training epochs is 50, and the batch size is 32. We also compare \texttt{Inter}-\texttt{BIN} with string edit distance and char n-gram based similarity comparison methods, respectively. We use the \texttt{python}-\texttt{Levenshtein} package \footnote{\url{https://pypi.org/project/python-Levenshtein/}} to calculate the edit distance of assembly basic block pairs. For char n-gram features, we use 4-gram and Jaccard similarity, which performs best in our evaluation. The AUC comparison results on Dataset1 are shown in \figurename \ref {bbcmp}.

From the figure we can see that, the neural network-based approach \texttt{INNEREYE} and \texttt{Inter}-\texttt{BIN} achieve good performance on basic block level comparison, significantly outperform \texttt{Gemini}'s statistical features, string edit distance, and char n-gram based methods. But \texttt{INNEREYE} does not introduce the inter-sequence interaction between different architectures. Redmond et al. use the linear mapping of position indexes to perform hard alignment of instruction pairs and construct cross-architecture contexts. However, this alignment way is not flexible and accurate. On basic blocks with similar semantics but significantly differ in instruction sequence length and order, it will generate a large number of incorrect instruction pair associations. Moreover, simply summing the instruction vectors will lose the internal sequence context information of the basic blocks, so the performance of their method is not ideal. \texttt{Inter}-\texttt{BIN}'s performance is superior to these approaches, proving that our automatic soft alignment of cross-architecture instruction pairs can effectively improve the binary comparison accuracy, and the bidirectional context encoder is suitable for instruction sequence representation. 

\subsubsection{Visualization}
\begin{figure*}[!t]
\centering
\quad
\subfigure{
\includegraphics[width=8cm]{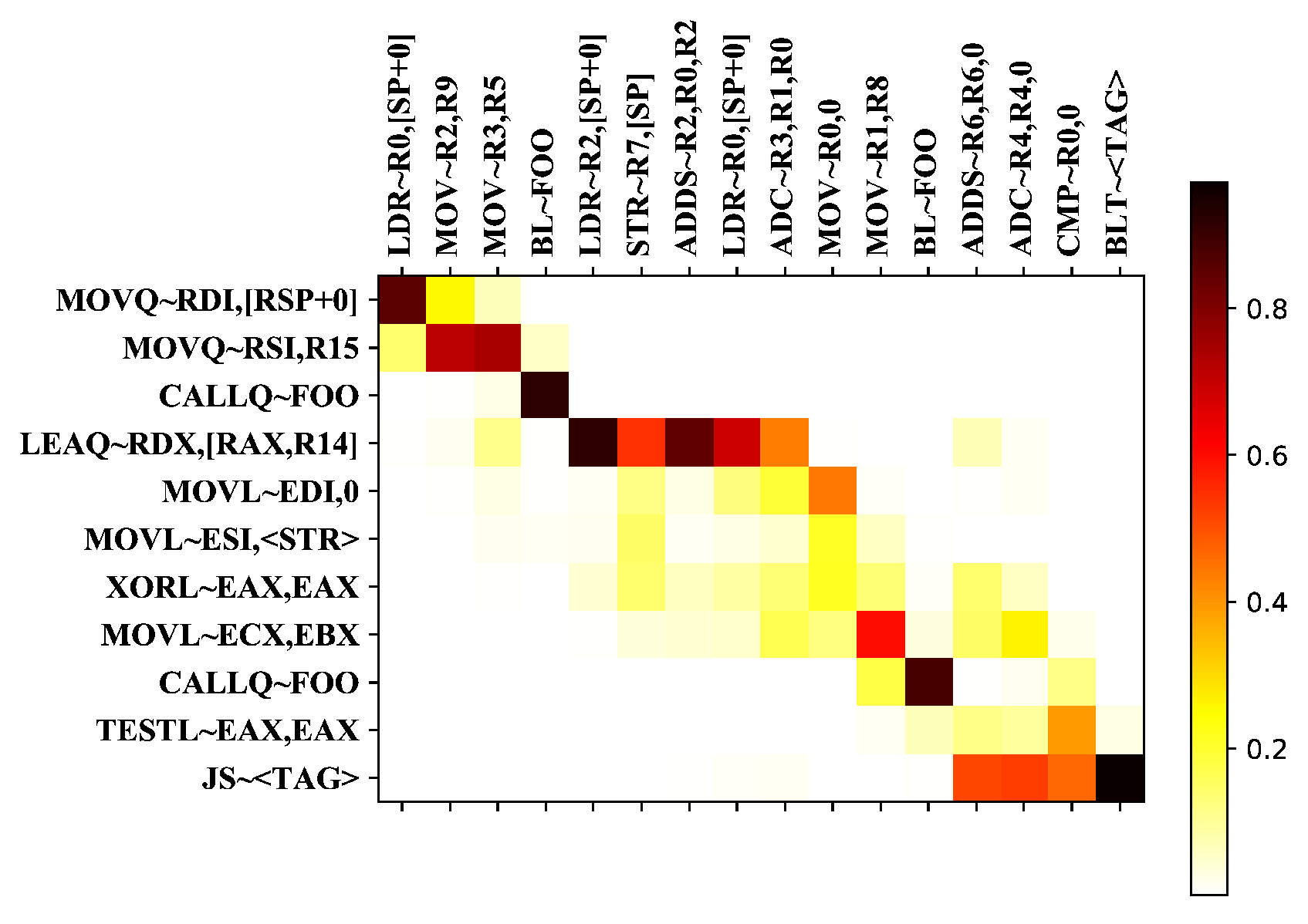}
}
\hspace{-2mm}
\quad
\subfigure{
\includegraphics[width=8cm]{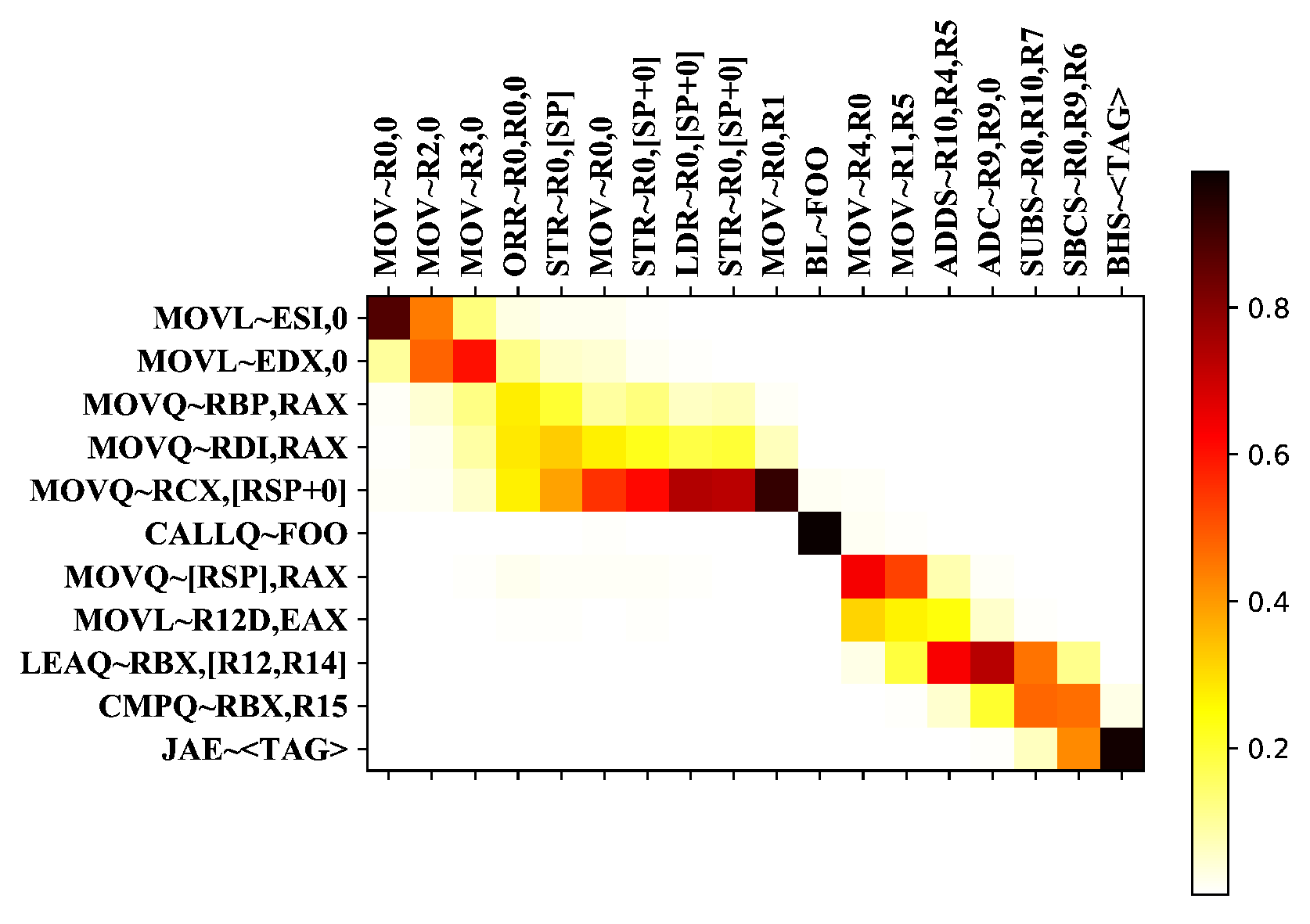}
}
\caption{Cross-architecture instruction alignment visualization of \texttt{Inter}-\texttt{BIN}.}
\label{visualization}
\vspace*{-0.5\baselineskip}
\end{figure*}

To present the automatic instructions soft alignment of \texttt{Inter}-\texttt{BIN}'s interaction module on instruction pairs across different architectures, we visualize the attention similarity matrices of the cross-architecture instruction sequences, as shown in \figurename \ref {visualization}. Darker colors indicate stronger semantic correlations between instruction pairs. 

From the figure, we can directly observe that the \emph{function call} instruction $CALL, FOO$ of the x86 architecture and $BL, FOO$ of the ARM architecture show high attention weight because they specific the same operation. Similarly, the \emph{jump if sign} instruction $JS, \textless{}TAG\textgreater{}$ of x86 and $B, \textless{}TAG\textgreater{}$ of ARM, the \emph{greater than or equal branch transfer} instructions $JAE, \textless{}TAG\textgreater{}$ and $BHS, \textless{}TAG\textgreater{}$, and the instructions related to memory access, including $MOVQ$, $MOVL$, $LEAQ$ of x86 and $MOV$, $LDR$, $STR$ of ARM have also been effectively linked. 

We perform in-depth analysis of the behavior achieved by the combination of multiple instructions. In the left subfigure, the first two instructions of x86 basic block use opcode $MOVQ$, general-purpose registers $RDI$ and $RSI$ and stack pointer register $RSP$ to realize function parameter transfer behavior. Meanwhile, the first three instructions of ARM basic block use opcode $LDR$ and $MOV$, general-purpose registers $R0$, $R2$, $R3$, and stack pointer $SP$ register to perform similar behavior. These cross-architecture instructions involved in the function parameter transfer behavior show higher similarity correlations in \figurename \ref {visualization}, and the subsequent function call instructions $CALL, FOO$ and $BL, FOO$ instructions are also associated with a high weight.

The visualization results show that \texttt{Inter}-\texttt{BIN} can achieve flexible and accurate soft alignment between semantically related instruction pairs of different architectures, which is significant for detecting cross-architecture binary code with similar functional semantics but different lexical expressions.

\subsubsection{Case study of Dataset1}

\textbf{Reduce false-positive cases.}
\texttt{INNEREYE} uses single directional LSTM to encode basic block level instruction sequences. The final basic block representations generated by this serialized modeling method are more affected by the last several instructions. Therefore, we observe that although the lengths of some dissimilar basic block pairs are very different, \texttt{INNEREYE} may incorrectly determine that they are similar. \texttt{Inter}-\texttt{BIN} can accurately determine these dissimilar pairs by the pairwise cross-architecture instructions alignment mechanism.

Table \ref {dissimilarcase} shows two of \texttt{INNEREYE}'s false-positive cases. For the first dissimilar pair, although the instruction length difference of the two basic blocks is 19, the instructions including $CALLQ$, $MOVQ$ opcodes at the end of the x86 basic block and the instructions including $BL$, $MOV$ at the end of the ARM basic block have high semantic similarities, so \texttt{INNEREYE} misjudges it as a similar pair. For the second dissimilar pair, the x86 basic block contains only a $JMP, \textless{}TAG\textgreater{}$ instruction, but \texttt{INNEREYE} determines it as a similar pair since the ARM basic block ending with the $B, \textless{}TAG\textgreater{}$ instruction. For these two examples, \texttt{Inter}-\texttt{BIN} can correctly determine that they are dissimilar basic block pairs. On the overall test set of Dataset1, \texttt{Inter}-\texttt{BIN} reduces the number of false-positive cases by 176 compared to \texttt{INNEREYE}.

\noindent \textbf{Reduce false-negative cases.}
At basic block level, some small pieces of binary code contain only a few instructions and provide a small amount of information. It is difficult for RNN-type models trained separately on different instruction sequences to identify these similar binary pairs accurately. Table \ref {similarcase} lists two similar basic block pairs that \texttt{INNEREYE} cannot correctly identify, while \texttt{Inter}-\texttt{BIN} with inter-sequence instruction interaction module can realize information transfer between the sequences and correctly identify them as similar.

\begin{table*}
\caption{Reduced false-positive cases: dissimilar pairs that are correctly classified by \texttt{Inter}-\texttt{BIN}, but misclassified by \texttt{INNEREYE}.}
\label{dissimilarcase}
\centering
\begin{tabular}{|c|c|c|c|}
	\hline
	\multicolumn{2}{|c|}{Dissimilar pair 1} & \multicolumn{2}{c|}{Dissimilar pair 2} \\
	\hline
	x86 & ARM & x86 & ARM \\ 
	\hline
    
    {\begin{tabular}[c]{@{}l@{}}
	MOVSLQ$\sim$RAX,{[}RIP+\textless{}TAG\textgreater{}{]}\\ 
	LEAL$\sim$ECX,{[}RAX+0{]} \\ 
	SHLQ$\sim$RAX,0  \\ 
	LEAQ$\sim$R15,{[}RAX+\textless{}TAG\textgreater{}{]}  \\ 
	ANDL$\sim$ECX,0  \\ 
	MOVL$\sim${[}RIP+\textless{}TAG\textgreater{}{]},ECX  \\ 
	LEAQ$\sim$R14,{[}RSP+0{]}\\      
	\\ ... 15 instructions\\ \\ 
	XORL$\sim$EAX,EAX \\ 
	MOVQ$\sim$RCX,R15  \\ 
	MOVQ$\sim$R15,{[}RSP+0{]}  \\ 
	CALLQ$\sim$FOO  \\ MOVQ$\sim$RDI,{[}RSP+0{]}\end{tabular}} & 
	
	{\begin{tabular}[c]{@{}l@{}}ADD$\sim$R0,R3,0 \\ 
	MOV$\sim$R1,R4 \\ 
	STR$\sim$R0,{[}R2{]}\\ 
	MOV$\sim$R0,R12\\ 
	BL$\sim$FOO\\ 
	MOV$\sim$R1,0\\ 
	STR$\sim$R4,{[}R0+0{]}\\ 
	STR$\sim$R1,{[}R0+0{]}\end{tabular}} & 
 	
 	{JMP$\sim$\textless{}TAG\textgreater{}} &
 	
 	{\begin{tabular}[c]{@{}l@{}}STR$\sim$R0,{[}R5{]}\\ 
 	LDR$\sim$R0,{[}SP+0{]}\\ 
 	STR$\sim$R9,{[}R0+0{]}   \\ 
 	MOV$\sim$R0,0\\ 
 	B$\sim$\textless{}TAG\textgreater{}\end{tabular}} \\
 	
    \hline
\end{tabular}
\end{table*}

\begin{table*}
\caption{Reduced false-negative cases: similar pairs that are correctly classified by \texttt{Inter}-\texttt{BIN}, but misclassified by \texttt{INNEREYE}.}
\label{similarcase}
\centering
\begin{tabular}{|c|c|c|c|}
	\hline
	\multicolumn{2}{|c|}{Similar pair 1} & \multicolumn{2}{c|}{Similar pair 2} \\
	\hline
	x86 & ARM & x86 & ARM \\ 
	\hline
    
    {\begin{tabular}[c]{@{}l@{}}MOVL$\sim$ESI,{[}RSI{]}\\
	CMPL$\sim$ESI,0\\ 
	JE$\sim$\textless{}TAG\textgreater{}\end{tabular}} & 
	
	{\begin{tabular}[c]{@{}l@{}}CMP$\sim$R1,0\\
	MVNNE$\sim$R0,0\\
	MOVNE$\sim$PC,LR\\
	B$\sim$\textless{}TAG\textgreater{}\end{tabular}} & 
 	
 	{\begin{tabular}[c]{@{}l@{}}
	MOVL$\sim${[}RIP+\textless{}TAG\textgreater{}{]},EBX\\ 
	MOVQ$\sim$R9,{[}RIP+\textless{}TAG\textgreater{}{]}\\
	MOVQ$\sim$RDI,{[}R9{]}\\
	MOVL$\sim${[}RDI+0{]},EBX\\ 
	XORL$\sim$R12D,R12D\\
	TESTL$\sim$EBX,EBX\\
	JNE$\sim$\textless{}TAG\textgreater \end{tabular}} &
 	
 	{\begin{tabular}[c]{@{}l@{}}LDR$\sim$R0,{[}R10+0{]}\\
	LDR$\sim$R1,{[}SP+0{]}\\
	BL$\sim$FOO\\
	CMP$\sim$R0,0\\
	BNE$\sim$\textless{}TAG\textgreater{}\end{tabular}} \\
 	
    \hline
\end{tabular}
\end{table*}

\subsection{Function level Evaluation}
In this section, we first show the specific information of the function level Dataset2, which is established by four different optimization levels. Then we evaluate \texttt{Inter}-\texttt{BIN} on Dataset2 and compare it with state-of-the-art approaches. Finally, we conduct a case analysis to explain why \texttt{Inter}-\texttt{BIN} can significantly improve the performance of the function level binary similarity comparison task.

\subsubsection{Comparison with State-of-the-Arts}
\begin{table}[]
\caption{Statistical information of Dataset2.}
\label{dataset2}
\centering
\renewcommand{\arraystretch}{1.1}
\begin{tabular}{|m{40pt}<{\centering}|m{35pt}<{\centering}|m{40pt}<{\centering}|m{35pt}<{\centering}|m{35pt}<{\centering}|}
		\hline
		Opt-level & \# Training pairs & \# Validation pairs & \# Testing pairs & Total \\
		\hline
		O0 & 88,050 & 10,877 & 9,784 & 108,711 \\
		O1 & 69,080 & 8,545 & 7,676 & 85,301 \\
		O2 & 60,611 & 7,495 & 6,735 & 74,841 \\
		O3 & 57,966 & 7,161 & 6,441 & 71,568 \\
		Cross-opts & 117,123 & 14,467 & 13,014 & 144,604 \\
		Total & 392,830 & 48,545 & 43,650 & 485,025 \\
		\hline
\end{tabular}
\end{table}

We divide Dataset2 into multiple subsets according to optimization levels to evaluate \texttt{Inter}-\texttt{BIN} on the function level binary matching task. Table \ref {dataset2} shows the statistics of Dataset2, containing a total of 485,025 function pairs. 

We compare \texttt{Inter}-\texttt{BIN} with \texttt{INNEREYE}, the best performing state-of-the-art approach in our basic block level experiments. Since we additionally use a new compiler \texttt{GCC}, the out-of-vocabulary (OOV) rate of \texttt{INNEREYE}'s pre-trained instruction embeddings on subsets of Dataset2 is in the range of 42.12\% to 53.11\%. So we replaced \texttt{INNEREYE}'s pre-training module with our multi-feature fusion-based instruction representation method as an improver variant.

We also make comparisons with a function level binary similarity comparison approach, \texttt{SAFE} \cite{massarelli2019SAFE}. Similar to \texttt{INNEREYE}, \texttt{SAFE} deploys an assembly instruction pre-training module $i2v$ and uses bi-directional RNN to encode function instruction sequences of different architectures separately. Then it performs self-attention on each sequence individually to enhance the role of instructions that are more important for the similarity matching result. Since \texttt{SAFE}'s instruction pre-training is generated under different compile settings and packages with our Dataset2, we cannot directly use the provided instruction embeddings. So we also replace its instruction embedding module $i2v$ with our own instruction representation module.

\begin{table*}[]
\caption{Function level cross-architecture binary similarity comparison results on Dataset2.}
\label{dataset2res}
\centering
\renewcommand{\arraystretch}{1.1}
\begin{tabular}{|m{40pt}<{\centering}|m{40pt}<{\centering}|m{40pt}<{\centering}|m{40pt}<{\centering}|m{45pt}<{\centering}|m{40pt}<{\centering}|m{40pt}<{\centering}|m{40pt}<{\centering}|m{45pt}<{\centering}|}

	\hline
	\multirow{2}{*}{Opt-level} & \multicolumn{4}{c|}{Precision@1} & \multicolumn{4}{c|}{MRR} \\ \cline{2-9} 
    & Pre-trained \texttt{INNEREYE} & Improved \texttt{INNEREYE} & Improved \texttt{SAFE} & \texttt{Inter}-\texttt{BIN} &  Pre-trained \texttt{INNEREYE} & Improved \texttt{INNEREYE}& Improved \texttt{SAFE} & \texttt{Inter}-\texttt{BIN} \\ [3pt]
    \hline
    O0  & 0.6486 & 0.7838 & 0.9266 & \textbf{0.9691}  & 0.7791  & 0.8742  & 0.9557 & \textbf{0.9833}  \\
    O1  & 0.6502 & 0.7199 & 0.8034 & \textbf{0.9214}  & 0.7614  & 0.8291 & 0.8810 & \textbf{0.9572}  \\
    
    O2  & 0.5014 & 0.6555 & 0.7255 & \textbf{0.9048}  & 0.6619  & 0.7819 & 0.8285 & \textbf{0.9433}  \\
    
    O3  & 0.6308 & 0.7419 & 0.7918 & \textbf{0.9179}  & 0.7647  & 0.8302 & 0.8723 & \textbf{0.9524}  \\
    
    Cross-opts  & 0.5951  & 0.6459 & 0.6967 & \textbf{0.8520}  & 0.7165 & 0.7695 & 0.8105 & \textbf{0.9108} \\
    \hline
\end{tabular}
\end{table*}

 We set up a ranking task on Dataset2, the number of negative samples per query function ($num\_neg$) is set to 20, and the number of training epochs is 300. The evaluation results are shown in Table \ref {dataset2res}. Benefit from the cross-architecture interaction module, \texttt{Inter}-\texttt{BIN} outperforms \texttt{INNEREYE} and \texttt{SAFE} by large margins at the function granularity. On the O0 optimization level subset, \texttt{Inter}-\texttt{BIN}'s precision@1 outperforms the improved \texttt{INNEREYE} variant by 18.53 points, and the MRR is increased by 0.2042. On the multiple optimization levels setting, \texttt{Inter}-\texttt{BIN}'s precision@1 outperform the improved variants of \texttt{INNEREYE} and \texttt{SAFE} by 20.61 points and 15.53 points. Moreover, our multi-feature fusion-based instruction representation module can effectively help \texttt{INNEREYE} avoid the performance loss caused by OOV. The precision@1 of \texttt{INNEREYE} is increased by 13.52 points on the O0 optimization level subset by extracting instruction char level spatial features and attributes of opcode and operands.

\subsubsection{Case study of Dataset2}
\begin{figure}[!t]
	\centering
	\includegraphics[width=8.5cm]{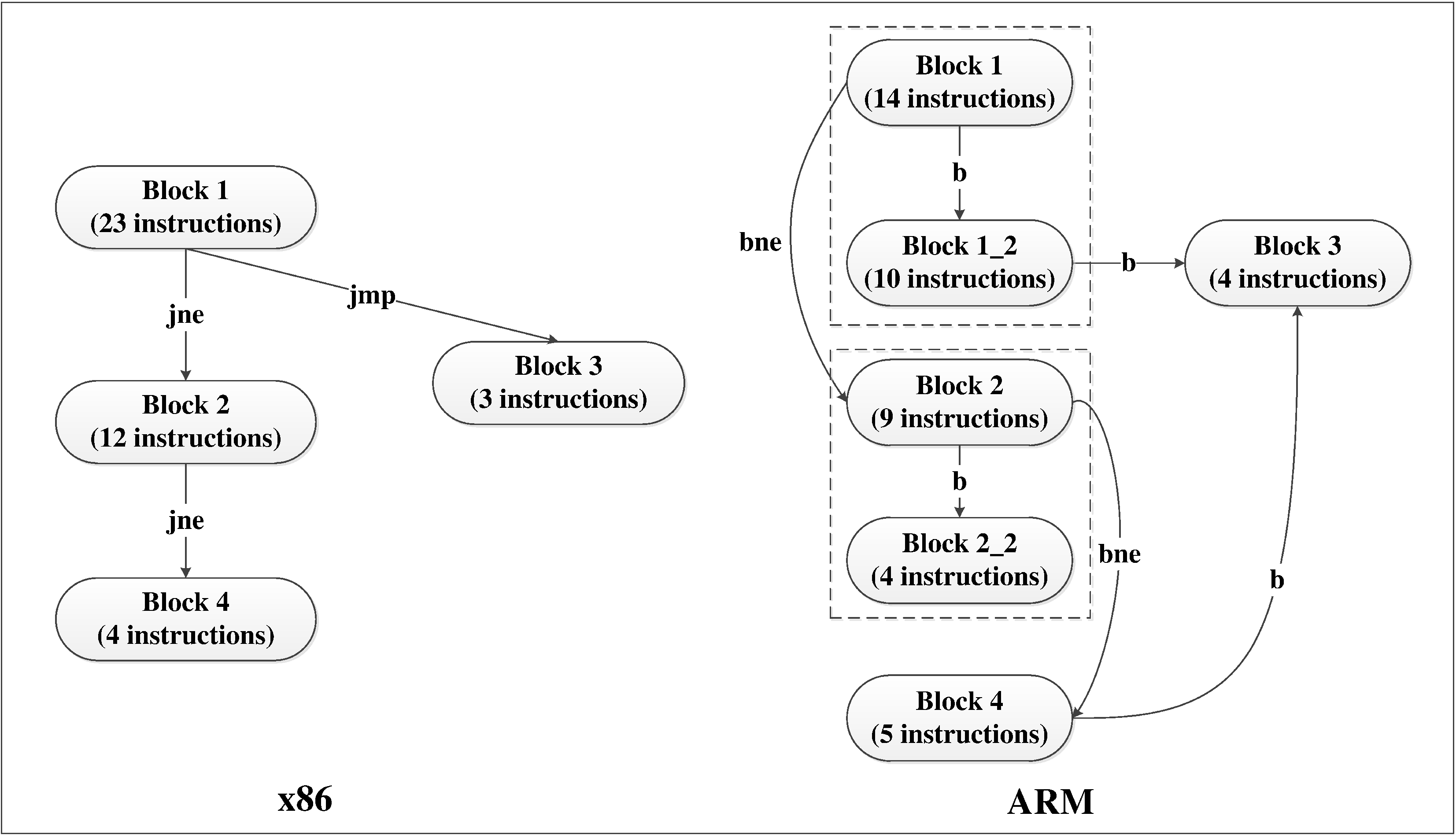}
	\caption{A similar function pair with changed CFGs that are successfully matched by \texttt{Inter}-\texttt{BIN}, but failed by \texttt{INNEREYE} and \texttt{SAFE}.}
	\label{funccase}
\end{figure}

In the function level binary similarity comparison scenario, we observe that the performance improvement of \texttt{Inter}-\texttt{BIN}'s for long function matching is more prominent. A long function usually contains multiple basic blocks and can be represented as a control-flow-graph (CFG) structure. Each node is a basic block, and the edges represent the transfer relationships between the blocks. \figurename \ref {funccase} shows the CFG structure of the disassembled $base\_name$ function from the \texttt{coreutils} 8.29 package. We omit the specific instruction sequences within the basic blocks. When the source code of this function is compiled on the x86 and ARM architectures, \emph{Block} 1 and \emph{Block} 2 are divided into two blocks, respectively, resulting in different CFG patterns.

\texttt{INNEREYE} and \texttt{SAFE} cannot match this function pair correctly, while \texttt{Inter}-\texttt{BIN} can rank the similarity of the correct cross-architecture candidate function at the top position. Although \texttt{Inter}-\texttt{BIN} models assembly instruction sequences by Bi-LSTM without directly using the structural information, the pairwise cross-architecture instruction alignment mechanism can handle changes of CFG patterns. Meanwhile, it can avoid the expensive CFG extraction and matching process.

\texttt{INNEREYE} also designs an \texttt{INNEREYE}-\texttt{CC} sub-system, which achieves the code component similarity matching by performing the longest common subsequence (LCS) algorithm on basic block sequences extracted from the CFG. However, \texttt{INNEREYE}-\texttt{CC} needs to train a large number of basic block embeddings to fully cover the fragments of target code components, which is not practical in real scenarios. 

\subsection{Evaluation on Real-world IoT Malware Dataset}
In this section, we use Dataset3, the cross-architecture IoT malware dataset \emph{CrossMal} collected in real network environments to evaluate \texttt{Inter}-\texttt{BIN}'s practicality and scalability. We first show the statistics information of \emph{CrossMal}. Then evaluate \texttt{Inter}-\texttt{BIN} and compare it with state-of-the-art approaches. We analyze specific malware cases to demonstrate the ability of \texttt{Inter}-\texttt{BIN} on cross-architecture reuse function detection. Finally, we evaluate the runtime overheads of \texttt{Inter}-\texttt{BIN} on the \emph{CrossMal} dataset.

\subsubsection{Cross-architecture Effectiveness Evaluation}
We construct \emph{CrossMal} as a function pairs collection to detect reused malware functions between IoT devices of different architectures, characterizing malicious behavior patterns from a finer granularity. Table \ref {dataset3} shows the overall information of \emph{CrossMal}, containing a total of 1,878,437 function pairs compiled from x86, ARM, and MIPS architectures \footnote{The \emph{CrossMal} dataset can be downloaded by the link \url{https://drive.google.com/file/d/1kluoLPojJ-gwyGHgu2uJ5kt_NDDVeITi/view?usp=sharing.}}. Through our statistics, most function's instruction sequence length in Dataset3 is in the range of 20 to 100. We use 50 as the threshold to divide Dataset3 into large-function subsets and small-function subsets to comprehensively evaluate \texttt{Inter}-\texttt{BIN}'s performance and runtime efficiency.

\begin{table}[]
	\caption{Statistical information of Dataset3.}
	\label{dataset3}
	\centering
	\renewcommand{\arraystretch}{1.1}
	\begin{tabular}{|m{45pt}<{\centering}|m{35pt}<{\centering}|m{40pt}<{\centering}|m{33pt}<{\centering}|m{35pt}<{\centering}|}
		\hline
		Settings & \# Training pairs & \# Validation pairs & \# Testing pairs & Total \\
		\hline
		x86-ARM & 409,588 & 45,510 & 42,754 & 497,852 \\
		x86-MIPS & 364,154 & 40,462 & 47,292 & 451,908 \\
		ARM-MIPS & 372,617 & 41,402 & 45,044 & 4590,63 \\
		Cross 3-arcs & 376,141 & 41,794 & 51,679 & 469,614 \\
		Total & 1,522,500 & 169,168 & 186,769 & 1,878,437 \\
		\hline
	\end{tabular}
\end{table}

We evaluate \texttt{Inter}-\texttt{BIN}'s function level similarity comparison performance on different settings of \emph{CrossMal} and make comparisons with state-of-the-art approaches. Since \emph{CrossMal} involves more architectures than the instruction pre-training code corpus used by \texttt{INNEREYE} and \texttt{SAFE}, we only compare \texttt{Inter}-\texttt{BIN} with the their variants improved by our multi-feature fusion-based instruction representation module to avoid the severe OOV problem. We set training epochs to 100, and other hyper-parameters settings are the same as section \uppercase\expandafter{\romannumeral8}.D.

\begin{table*}[]
\centering
\renewcommand{\arraystretch}{1.1}
\caption{Cross-architecture binary similarity comparison results on Dataset3.}
\label{dataset3res}
\begin{threeparttable}
	
\begin{tabular}{|m{38pt}<{\centering}|m{38pt}<{\centering}|m{35pt}<{\centering}|m{30pt}<{\centering}|m{45pt}<{\centering}|m{35pt}<{\centering}|m{30pt}<{\centering}|m{45pt}<{\centering}|m{35pt}<{\centering}|m{30pt}<{\centering}|m{45pt}<{\centering}|}
\hline
\multicolumn{2}{|c|}{\multirow{2}{*}{Settings}} & \multicolumn{3}{c|}{All-functions}        & \multicolumn{3}{c|}{Large-functions}      & \multicolumn{3}{c|}{Small-functions}      \\ \cline{3-11} 
\multicolumn{2}{|c|}{}                          & \texttt{INNEREYE} & \texttt{SAFE} & \texttt{Inter}-\texttt{BIN} & \texttt{INNEREYE} & \texttt{SAFE} & \texttt{Inter}-\texttt{BIN} & \texttt{INNEREYE} & \texttt{SAFE} & \texttt{Inter}-\texttt{BIN} \\ \hline
\multirow{2}{*}{x86-ARM}        & Precision@1   
& 0.8922  & 0.8994 & \textbf{0.9025} & 0.9386  & 0.9176 & \textbf{0.9582}  & 0.8315 & 0.8525 & \textbf{0.8734} \\ \cline{2-11}					& MRR           
& 0.9398  & 0.9440 & \textbf{0.9468} & 0.9674  & 0.9563 & \textbf{0.9789}  & 0.9036 & 0.9170 & \textbf{0.9299} \\                           
\hline

\multirow{2}{*}{x86-MIPS}       & Precision@1   
& 0.9067  & 0.9143 & \textbf{0.9172} & 0.8760  & 0.9058 & \textbf{0.9428}  & 0.8656 & 0.8920 &  \textbf{0.8992} \\ \cline{2-11}                     & MRR           
& 0.9502  & 0.9547 & \textbf{0.9571} & 0.9309  & 0.9514 & \textbf{0.9708}  & 0.9283 & 0.9419 &  \textbf{0.9467} \\                                
\hline

\multirow{2}{*}{ARM-MIPS}       & Precision@1   
& 0.8826  & 0.8155 & \textbf{0.9044} & 0.9129  & 0.8946 & \textbf{0.9382}  & 0.8376 & 0.8014 &  \textbf{0.8679} \\ \cline{2-11}                    & MRR           
& 0.9375  & 0.9001 & \textbf{0.9495} & 0.9539  & 0.9458 & \textbf{0.9683}  & 0.9118 & 0.8915 &  \textbf{0.9285} \\                                 
\hline

\multirow{2}{*}{Cross 3-arcs}   & Precision@1   
& 0.8674  & 0.8781 & \textbf{0.9010} & 0.8135  & 0.8261 & \textbf{0.9260}  & 0.8594 & 0.8454 &  \textbf{0.8895} \\ \cline{2-11}                    & MRR           
& 0.9277  & 0.9339 & \textbf{0.9464} & 0.8942  & 0.9066 & \textbf{0.9624}  & 0.9208 & 0.9110 &  \textbf{0.9381} \\ 
\hline
\end{tabular}
	\begin{tablenotes}
	\footnotesize
	\item[*]  \texttt{INNEREYE} and \texttt{SAFE} refer to the variants improved by our \emph{multi-feature fusion-based instruction representation module}.
\end{tablenotes}
\end{threeparttable}
\vspace*{-1.2\baselineskip}
\end{table*}

Table \ref {dataset3res} shows that on different settings and the subsets of different function scales, \texttt{Inter}-\texttt{BIN} all outperforms the state-of-the-art approaches without the cross-architecture interaction module. On the test set containing 42,285 function pairs from three CPU architectures, \texttt{Inter}-\texttt{BIN}'s precision@1 reaches 0.9010, and MRR is up to 0.9464. Through further analysis, we found that the larger function contains more semantic information for fully interacting with another instruction sequence, so the improvement of \texttt{Inter}-\texttt{BIN}'s cross-architecture interaction module is significant. On the large-function subset of three mixture architectures, the precision@1 \texttt{Inter}-\texttt{BIN} are 11.25 points and 9.99 points higher than the multi-feature fusion-based \texttt{INNEREYE} and \texttt{SAFE}. 

\subsubsection{Case study of CrossMal}
\emph{CrossMal} contains malware targeting IoT devices of different architectures captured by the \texttt{IoTCMal} honeypot \cite{wang2020iotcmal}. We manually analyze two binary files named \emph{cc9x86} and \emph{cc9arm6} of the IoT malware family Gafgyt \footnote{The sha256 values of these two binaries are \emph{af47b7f0b887d8ce09a3a260945b658dc9b5323c5f0efa2e66c67905d0c0dbe3} and \emph{7c0e22da32c8ce46927e3f7671535d6f75d6bcdcf70a1919afe1695dcd1c2c33}.}. They were complied on x86 and ARM architectures and caught by us on August 2, 2020. Their code structures are highly similar and share a large number of reuse functions. In all the cross-architecture function pairs implemented by malicious developers, \texttt{INNEREYE} cannot correctly match the \emph{sendSTD}, \emph{sendVSE}, \emph{makeIPPacket}, \emph{makeVSEPacket}, and \emph{connectTimeout} functions compiled on the x86 and ARM platforms, while \texttt{Inter}-\texttt{BIN} can match these cross-architecture function pairs correctly. We regard ranking the candidate function compiled from the same source code at the top as a successful match. Due to the cross-architecture interaction module, the overall successful matching rate of the reuse function pairs of these two binary files increased from 79\% to 85\%. Among the 60 industrial anti-virus scanners on VirusTotal, six can only detect one of the binary files, but not the other, and 29 can detect neither of them. 

When using only the user-defined \emph{main} function developed by malicious developers for cross-architecture binary comparison, \texttt{Inter}-\texttt{BIN} can achieve a matching accuracy of 95.16\% on the dataset of of ARM and MIPS architectures which are widely used in IoT devices. As a reference, the accuracy of \texttt{INNEREYE} and \texttt{SAFE} are 86.29\% and 91.94\%, respectively. In conclusion, the cross-architecture interaction module introduced by \texttt{Inter}-\texttt{BIN} enhances the reuse function detection accuracy, which is meaningful for preventing the rapid spread of IoT malware on devices of different architectures.

\subsubsection{Runtime Overhead on CrossMal}
\begin{figure*}[t]
	\centering
	\quad
	\subfigure[Preprocessing time]{
		\includegraphics[width=5.8cm]{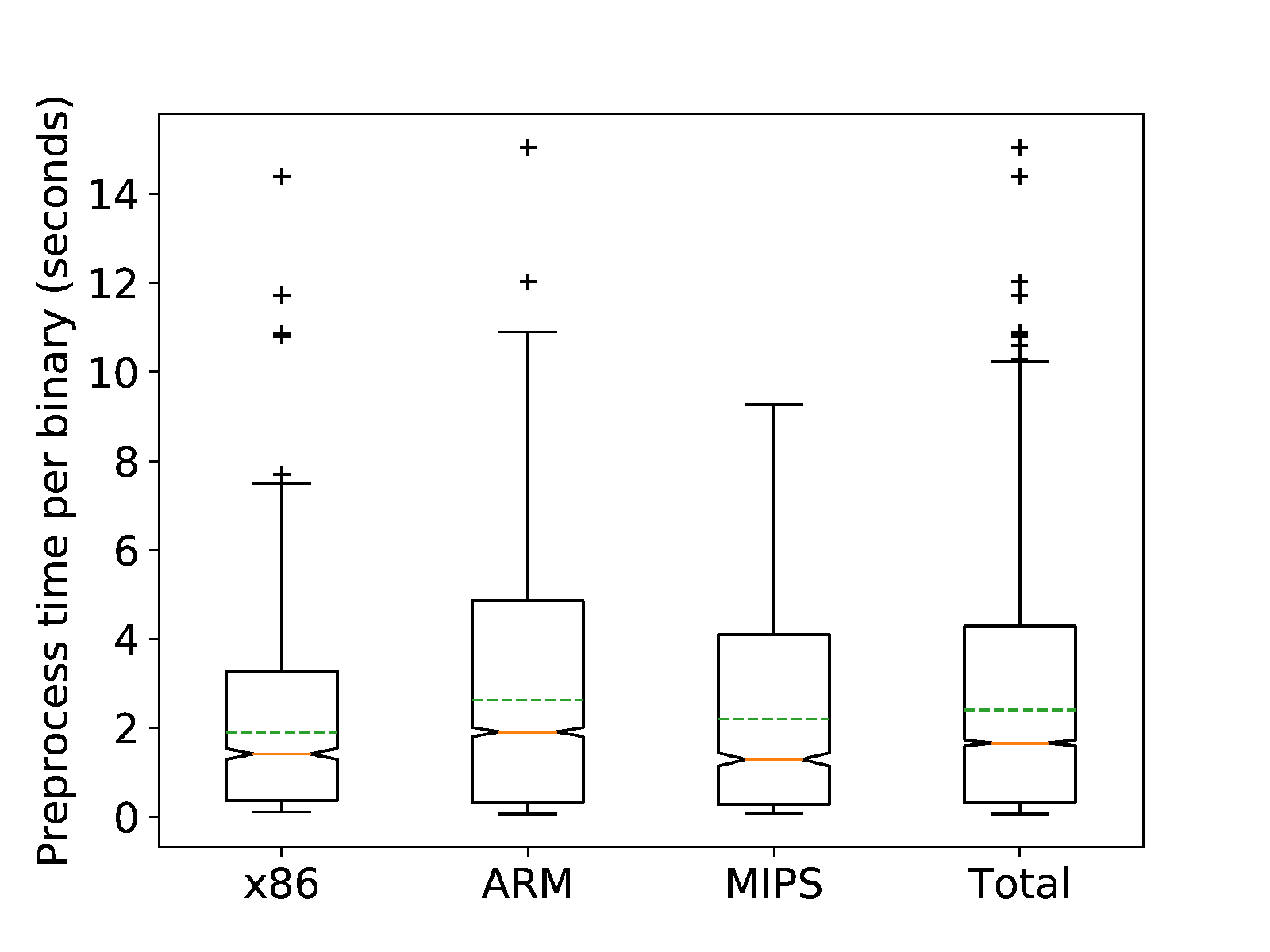}
	}
	\hspace{-10mm}
	\quad
	\subfigure[Off-line training time]{
		\includegraphics[width=6cm]{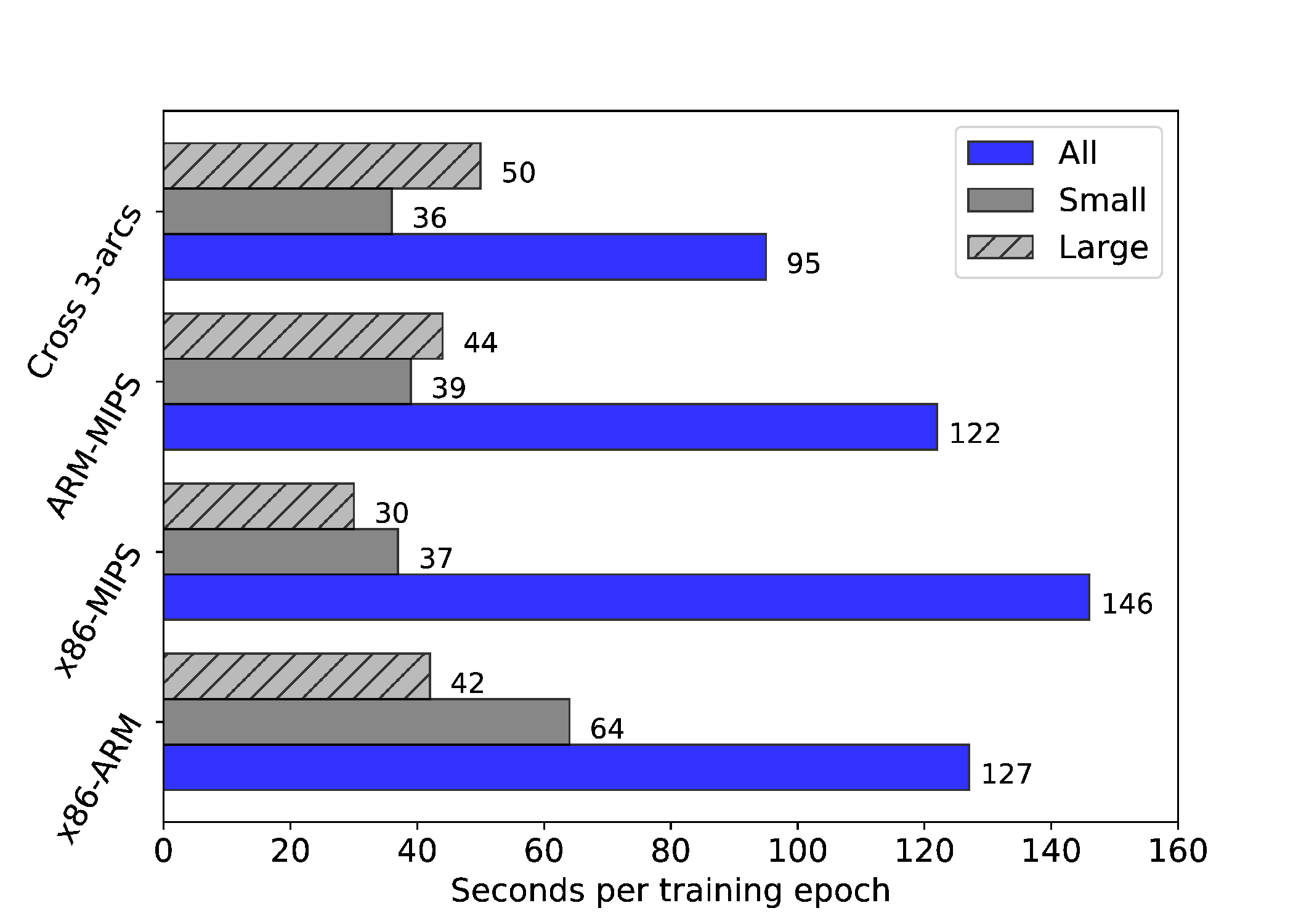}
		
	}
	\hspace{-10mm}
	\quad
	\subfigure[On-line prediction time]{
		\includegraphics[width=6cm]{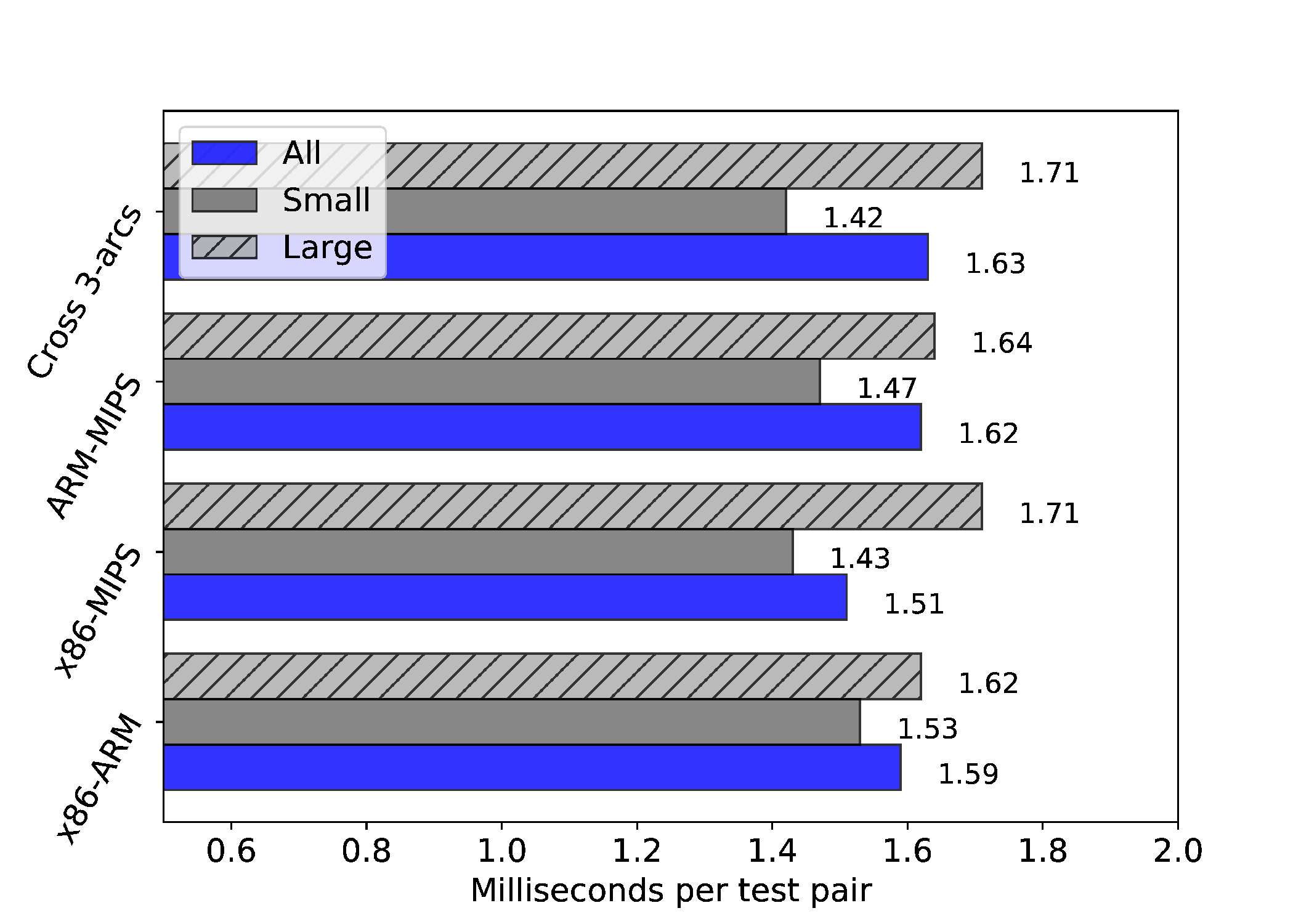}
	}
	\caption{Runtime overheads on Dataset3}
	\label{runtime}
	\vspace*{-1.2\baselineskip}
\end{figure*}

\figurename \ref{runtime}. (a) to \figurename \ref{runtime}. (c) respectively show the runtime overheads of \texttt{Inter}-\texttt{BIN} on four cross-architecture settings and three function scales for binary preprocessing, off-line training, and on-line prediction. And we compare the training and prediction time of \texttt{INNEREYE} and \texttt{SAFE} variants improved by our multi-feature fusion-based instruction representation module. Our evaluations are performed on a server with four 8-core Intel Xeon Silver-4110 CPUs running at 2.10GHz and 128GB of physical memory. The deep neural network of \texttt{Inter}-\texttt{BIN} runs on a GeForce RTX 2080 graphic card.
\\ \textbf{Preprocessing time.} We utilize multi-core CPUs to run 20 pre-processing procedures in parallel and skip binaries that take more than 120 seconds to be successfully processed. For most instances, the function-level instruction sequence extraction performed by radare2 \footnote{\url{https://www.radare.org/}} can complete within 5 seconds.
\\ \textbf{Off-line training time.} \texttt{Inter}-\texttt{BIN}'s model training time is positively correlated with the size of the train sets and the number of epochs. As shown in \figurename \ref{runtime}. (b), each bar represents the runtime for all samples in the train set completing one epoch of training. On the dataset across three architectures, \texttt{Inter}-\texttt{BIN} takes 123.95 seconds to complete an epoch of training on 380,564 function pairs, and can complete 100 training epochs within 3.5 hours.
\\ \textbf{On-line prediction time.} \texttt{Inter}-\texttt{BIN} can achieve efficient on-line similarity predictions of unknown cross-architecture binary snippets pairs. For the dataset across three architectures, \texttt{Inter}-\texttt{BIN} can return the semantic similarity matching result for a large query function within 1.63 milliseconds. Under all settings, \texttt{Inter}-\texttt{BIN} can predict the comparison result of the query and candidate function pair within two milliseconds. 

The cross-architecture semantic interactions makes the model training time of \texttt{Inter}-\texttt{BIN} slightly longer than \texttt{INNEREYE}, but we avoid the laborious instruction pre-training on the large-scale external code corpus. \texttt{INNEREYE} processed over 6,115K basic blocks only for the x86 platform, and \texttt{SAFE} uses 1,299K unique functions containing 190 million assembly code lines for two architectures instruction pre-training, which is undoubtedly a heavyweight work. The overall off-line training speed of \texttt{Inter}-\texttt{BIN} and \texttt{INNEREYE} is significantly faster than \texttt{SAFE}, and the on-line prediction time of the three approaches is in similar ranges. In conclusion, \texttt{Inter}-\texttt{BIN} can achieve efficient cross-architecture function similarity comparison on large-scale real-world IoT malware collection.

\section{Related Work}
\subsection{Binary Similarity Comparison}
Binary similarity comparison has a wide range of applications in software security areas, such as patch analysis, bug search, and code clone detection. Previous binary similarity comparison approaches can be systematically divided into traditional methods and learning-based methods.

\noindent \textbf{Traditional methods.} Traditional binary matching approaches are usually implemented by program static analysis \cite{pewny2015cross} \cite{xu2017spain} \cite{david2016statistical} \cite{chandramohan2016bingo} \cite{gao2008binhunt} \cite{eschweiler2016discovre} and dynamic analysis \cite{egele2014blanket} \cite{ming2012ibinhunt}  \cite{tian2013dkisb} \cite{jiang2020similarity} techniques. \texttt{Esh} \cite{david2016statistical} used a theorem prover to measure the semantic similarity of decomposed small code fragments. \texttt{ImOpt} \cite{jiang2020similarity} adopted the SSA (static single-assignment) transforming algorithm for code re-optimizing, improving the binary matching accuracy across different optimization levels. Most of these approaches only support binary similarity comparison under a single architecture. \texttt{Multi}-\texttt{MH} \cite{pewny2015cross} is the first cross-architecture binary matching method. It converted the binaries of different CPU architectures to intermediate representation (IR) code and then performed function indexing according to the input and output semantics of basic blocks. However, It is computationally expensive and unscalable on large-scale binary collections.

\noindent \textbf{Learning-based methods.} To improve the code analysis accuracy and efficiency, researchers have recently commenced using learning-based binary similarity matching approaches \cite{zuo2018neural} \cite{redmond2018cross} \cite{duan2020deepbindiff} \cite{massarelli2019SAFE}  \cite{xu2017neural} \cite{ding2019asm2vec} \cite{gao2018vulseeker} \cite{yu2020order} \cite{liu2018alphadiff}. \texttt{Asm2Vec} \cite{ding2019asm2vec} used the PV-DM model on instruction execution traces to train function embeddings. \texttt{DEEPBINDIFF} \cite{duan2020deepbindiff} generated structural basic block embeddings by TADW algorithm, and combined them with context-based embeddings. These methods are designed for binary matching across versions or optimization levels, without supporting different architectures. For cross-architecture scenarios, \texttt{Gemini} used \cite{xu2017neural} a Structure2vec network and the cosine similarity to achieve control-flow-graphs (CFG) comparison. \texttt{VulSeeker} \cite{gao2018vulseeker} extended \texttt{Gemini}'s graph structure by adding data flow edges. However, extracting accurate CFG is a non-trivial job relying on complex program control flow analysis techniques. \texttt{INNEREYE} \cite{zuo2018neural} deployed skip-gram model for instruction pre-training and used LSTM to model instruction sequences separately. Redmond et al. \cite{redmond2018cross} designed a rough position-based hard alignment method to perform association of cross-architecture instructions. In addition, many precious methods rely on the large-scale external code corpus to pre-train instructions \cite{zuo2018neural} \cite{redmond2018cross} \cite{duan2020deepbindiff} \cite{yu2020order}, which is labor-intensive and prone to suffer the out-of-vocabulary (OOV) problem. The \texttt{Inter}-\texttt{BIN} system we propose deploys a multi-feature fusion-based instruction representation module to avoid OOV, and we design an inter-sequence interaction mechanism to perform automatic soft alignment of cross-architecture instructions.

\subsection{IoT Malware Detection}
Costin et al. \cite{costin2018iot} manually collected unique resources of over 60 IoT malware families and pointed out that the current security community is still inadequate in vulnerability management and malware defense solutions. Alasmary et al. \cite{alasmary2019analyzing} extracted the graph-theoretic features of CFG and established a deep learning-based IoT malware detection model. \texttt{MSimDroid} \cite{wu2020detection} proposed a multi-dimensional similarity-based method to detect fake IoTs app in the markets. 
\texttt{IoTPOT} \cite{pa2016iotpot} and \texttt{IoTCMal} \cite{wang2020iotcmal} can simulate fragile IoT devices in the public network and capture attacks targeting them. The analysis of captured samples showed that some IoT malware families evolved rapidly in a short period and disseminated malware on a large number of devices of different CPU architectures. Lee et al. \cite{lee2020cross} extracted statistical features of printable strings to characterize IoT malware of different architectures. However, string-based features are not robust enough and can easily be modified or obfuscated by malicious developers. We design an interaction-based semantic similarity comparison method for binary assembly instruction sequences, which can detect reuse IoT malware spread on devices of different architectures.

\subsection{Deep Sequence Matching Models}
The design of \texttt{Inter}-\texttt{BIN} is inspired by the text sequence matching technique in natural language processing (NLP). Existing text sequence matching models can be divided into two categories: sequence encoding models \cite{conneau2017supervised} \cite{nie2017shortcut} and sequence pair interaction models \cite{pang2016text} \cite{wang2016machine} \cite{chen2016enhanced} \cite{lan2018neural}. Infersent \cite{conneau2017supervised} trained universal sentence representations and perform evaluations on 12 transfer tasks. \texttt{MatchPyramid} \cite{pang2016text} generated a corresponding matching matrix based on different word-level similarity metrics. \texttt{MatchLSTM} \cite{wang2016machine} designed a hypothesis to premise attention to realize semantic interaction on textual entailment task. \texttt{ESIM} \cite{chen2016enhanced} designed an enhanced natural language inference model considering recursive architectures in both local inference modeling and inference composition. Inspired by interaction-based text matching methods, we design an automatic soft alignment mechanism of inter-sequence instruction pairs to improve the cross-architecture binary matching accuracy.

\section{Conclusion and Future Work}
This paper proposes the first use of a \emph{deep neural network} with an \emph{interaction mechanism} for cross-architecture IoT binary similarity comparison, and provides effective security solutions against IoT malware threats. To avoid the heavy workload and the OOV problem of commonly used instruction pre-training approaches, we design a multi-feature fusion-based instruction representation method to extract the spatial and semantic features of assembly instructions. To overcome the lexical and syntax variations of similar binaries from different architectures, we perform inter-binary semantic interaction by co-attention, which can realize automatic soft alignment of assembly instruction pairs.

We implement our solution as an end-to-end multi-granularity cross-architecture binary similarity comparison system, \texttt{Inter}-\texttt{BIN}. Experimental results show that \texttt{Inter}-\texttt{BIN} outperforms state-of-the-art approaches on both basic block level and function level inputs. We establish \emph{CrossMal}, a large-scale IoT malware dataset containing 1,878,437 cross-architecture function pairs. Experiments and case analysis on \emph{CrossMal} prove that \texttt{Inter}-\texttt{BIN} is practical and scalable in real-world reuse function detection scenarios, which is significant for defending the IoT devices against malware that disseminates rapidly across different architectures.

In the future, we will further study the performance of our designed cross-architecture instructions alignment mechanism on the code containment problem, which determines whether a query piece of code is contained in another code snippet of a different architecture. It can help discover small malicious payloads injected into benign code modules. In addition, \texttt{Inter}-\texttt{BIN}'s function level instruction sequence encoding module follows the state-of-the-art approach \texttt{SAFE} \cite{massarelli2019SAFE}, which directly treated the function level assembly instructions as a sequence. In the future, we consider sampling the possible execution paths of functions within the control-flow-graph to explore whether it can perform semantic modeling better. 

\ifCLASSOPTIONcaptionsoff
  \newpage
\fi

\scriptsize
\bibliographystyle{./IEEEtran}
\bibliography{./final_version}

\end{document}